\documentclass[twocolumn,journal]{IEEEtran}
\usepackage[T1]{fontenc}
\usepackage[latin9]{inputenc}
\usepackage{amsbsy}
\usepackage{amssymb}
\usepackage{graphicx}
\usepackage[unicode=true,
 bookmarks=true,bookmarksnumbered=true,bookmarksopen=true,bookmarksopenlevel=1,
 breaklinks=false,pdfborder={0 0 0},pdfborderstyle={},backref=false,colorlinks=false]
 {hyperref}
\hypersetup{pdftitle={Exceptional Points in Gyrator-Based Circuit and Nonlinear High-Sensitivity Oscillator},
 pdfauthor={Rouhi et al.},
 pdfpagelayout=OneColumn, pdfnewwindow=true, pdfstartview=XYZ, plainpages=false}

\usepackage{fancyhdr}

\makeatletter
\let\oldforeign@language\foreign@language
\DeclareRobustCommand{\foreign@language}[1]{%
  \lowercase{\oldforeign@language{#1}}}

\usepackage[caption=false,font=footnotesize]{subfig}

\makeatother

\begin{document}
\title{Exceptional Points in Gyrator-Based Circuit and Nonlinear High-Sensitivity Oscillator}
\author{Alireza Nikzamir, Kasra Rouhi, Alexander Figotin, and Filippo Capolino\thanks{Alireza Nikzamir, Kasra Rouhi, and Filippo Capolino are with the Department
of Electrical Engineering and Computer Science, University of California,
Irvine, CA 92697 USA, e-mails: \protect\href{mailto:anikzami@uci.edu}{anikzami@uci.edu},
\protect\href{mailto:kasra.rouhi@uci.edu}{kasra.rouhi@uci.edu}, and
\protect\href{mailto:f.capolino@uci.edu}{f.capolino@uci.edu}.}\thanks{Alexander Figotin is with the Department of Mathematics, University
of California, Irvine, CA 92697 USA, e-mail: \protect\href{mailto:afigotin@uci.edu}{afigotin@uci.edu}.}}
\markboth{}{Alireza Nikzamir \MakeLowercase{\emph{et al.}}: Your Title}
\maketitle

\thispagestyle{fancy}

\begin{abstract}
We present a scheme for high-sensitive oscillators based on an exceptional
point of degeneracy (EPD) in a circuit made of two LC resonators coupled
by a gyrator. The frequency of oscillation is very sensitive to perturbations
of a circuit element, like a capacitor. We show conditions that lead
to an EPD, assuming one of the two resonators is composed of an inductor
and a capacitor of negative values. The EPD occurrence and sensitivity
to perturbations in the linear case are demonstrated by showing that
the eigenfrequency bifurcation around the EPD is described by the
relevant Puiseux (fractional power) series expansion. We also investigate
the effect of small losses in the system and show that they lead to
instability. We fabricate the circuit, and exploit its instability
and nonlinearity, observing experimentally stable self-oscillations
under the saturated regime. We measure the circuit's sensitivity to
a small capacitor perturbation. A shift in frequency of oscillation
after saturation is well detectable with very distinct spectral peaks
with 10 Hz linewidth, clean until -70 dB from the peak value. The
sensitivity is (i) higher than the one of a comparable simple LC linear
resonator, (ii) comparable or better than other published EPD circuits,
and (iii) applicable to both negative and positive values of the capacitance
perturbation, contrary to what happens in PT-symmetric circuits. The
proposed scheme can pave the way for a new generation of high-sensitive
sensors to measure slight variations in physical, chemical or biological
quantities.
\end{abstract}


\IEEEpeerreviewmaketitle{}

\section{Introduction}

\IEEEPARstart{R}{ecent} advancements associated with the concept
of exceptional points of degeneracy (EPDs) have attracted a surge
of interest due to their potential for various applications. An EPD
is a point in the parameter space of a system for which the eigenvalues
and the eigenvectors of the relevant matrix coalesce \cite{Vishik_1960_Solution,Lancaster1964On,Kato1966Perturbation,Heiss1990Avoided,Seyranian1993Sensitivity,Bender2002Generalized,Heiss2004Exceptional,Seyranian2005Coupling,Heiss2012Thephysics}.
The EPD concept has been investigated in temporally periodic electric
and mechanical systems \cite{kazemi2022experimental,Nikzamir2023Time},
in coupled-resonator systems with loss and/or gain under parity-time
symmetry \cite{Bender1998Real,Heiss2004Exceptional,Schindler2011Experimental,Kazemi2022High}.
The EPD concept using saturable nonlinear gain has been exploited
in conceiving oscillators based on two coupled transmission line \cite{Kazemi2022High,Moncada2024Frequency-Domain}
and two resonator circuits \cite{Nikzamir2022Highly}. Since the characterizing
feature of an exceptional point is the full degeneracy of at least
two eigenmodes, as mentioned in \cite{berry2004Physics}, the significance
of referring to it as \textquotedblleft degeneracy\textquotedblright{}
is here emphasized, hence including the D in EPD. In essence, an EPD
is obtained when the system matrix associated to a linear system is
similar to a matrix that comprises a non-trivial Jordan block. In
recent years, frequency splitting phenomena at EPDs have been proposed
for sensing applications \cite{wiersig2014enhancing,Rouhi2021exceptional,rouhi2024simple}.
Frequency splitting occurs at degenerate resonance frequencies where
system eigenmodes coalesce. Such a degenerate resonance frequency
is extremely sensitive to a small perturbation in system parameters.
This perturbation leads to a shift in the system resonance frequency
that can be detected and measured. This concept has been exploited
in new sensing schemes such as optical microcavities \cite{Chen2017Exceptional},
optical gyroscopes \cite{chow1985ring,sunada2007design} and mass
sensor \cite{djorwe2019exceptional}. Recently, EPD in nonlinear systems
has gained interest by showing its potential in advancing sensing
technologies and stability analysis. These studies \cite{suntharalingam2023noise,Bai2024Observation}
illustrate how nonlinear dynamics at EPDs enhance sensor sensitivity
and signal-to-noise ratio and pave the way for innovative electronic
system designs, underlining EPDs' critical role in sensing applications.

It has been recently shown that negative capacitors and inductors
are useful to realize EPDs in a system made of two resonators coupled
via a gyrator \cite{figotin2020synthesis,Figotin2021Perturbations,rouhi2022exceptional,rouhi2022high}.
These non-passive negative reactive components are synthesized with
negative impedance converters (NICs) or negative impedance inverters
(NIIs), which produce a negative capacitor or a negative inductor
with feedback circuits \cite{White2012Variable}. Negative capacitances
and inductances are largely used in electronics where negative capacitors
are obtained with op amps \cite{Kopp1965Negative,Horowitz1989art}
or with other semiconductor devices \cite{Ershov1998Negative}. Negative
inductances were obtained as early as 1965 using a grounded NII \cite{Kopp1965Negative},
and various circuits have been proposed for floating negative inductance
using different types of op amps for operation below 1 MHz. An ideal
gyrator is a two-port network that transforms a current into a voltage
and vice versa and causes 180 degrees phase shift difference in the
signal transmission from one side to the other \cite{tellegen1948gyrator}.
Gyrators have been designed using operational amplifiers (op amps)
\cite{Inigo1971Gyrator} or microwave circuits \cite{Hogan1952Ferromagnetic}.

In this paper, we explore for the first time the saturation regime
due to the nonlinearity in active negative inductance and negative
capacitance, in an EPD resonator topology based on a gyrator, and
explore the measured high sensitivity. We describe several EPD features
in gyrator-based coupled resonant circuits, where two LC resonators
(series-series and parallel-parallel configurations) are coupled to
each other through a gyrator. We illustrate the necessary conditions
to obtain the EPD in both parallel and series resonant circuit configurations
and show the signal behavior using time domain simulations. We also
provide a frequency domain analysis in terms of phasors and show that
the EPD corresponds to a double zero of the total impedance defining
the resonance. Importantly, we discuss the effect of additional losses
in the system and show how they make the circuit unstable. The effect
of the circuit's nonlinearities is observed experimentally after saturation
is achieved, leading to a stable oscillatory regime. When the system
is perturbed away from its EPD, the self-oscillation frequency is
shifted, and such a shift is measured to determine the circuit's sensitivity.
Compared to our previous studies in \cite{figotin2020synthesis,Figotin2021Perturbations,rouhi2022exceptional,rouhi2022high},
we focus on the analysis of the series-series configuration including
losses that we did not explore before; we also analyze a parallel-parallel
configuration, and fabricate a gyrator-based circuit for the first
time. We then observe experimentally the self oscillatory regime under
saturation and perturb a capacitance value to measure the oscillation
frequency shifts. In addition, we compare the sensitivity of our proposed
circuit to previous linear and nonlinear circuits supporting EPD \cite{Schindler2011Experimental,Nikzamir2022Highly,kazemi2022experimental,Moncada2024Frequency-Domain},
highlighting how its sensitivity is comparable and emphasizing the
capability of detecting small perturbations. The proposed circuit
and method have promising applications in ultrasensitive sensors at
various operating frequencies.

\section{EPD in Parallel Configuration}

We show a configuration in which we get an EPD by using a gyrator-based
circuit. Two parallel resonators are utilized in two different lossy/lossless
configurations. We briefly introduce the gyrator element and later
on, we write the required circuit equations in the Liouvillian formalism.
Then, we solve the eigenvalue problem to calculate the resonant frequencies
(i.e., the eigenfrequencies) and determine the conditions for obtaining
EPD at a desired frequency in a lossy circuit. We discuss the conditions
for real-valued EPD frequency and stability in the system. In order
to provide a comprehensive analysis of the presented circuit and its
stability, in Section \ref{subsec:PP}, we study the eigenfrequencies
in the lossless resonators, and we verify our theoretical calculations
by using a time-domain circuit simulator (Keysight Advanced Design
System (ADS)). Then, we provide an example and the eigenfrequency
dispersion with respect to changes in parameters and we show the perturbation
effects on the circuit's eigenfrequencies.

\subsection{EPD in Lossy Parallel Circuit\label{subsec:Parallel-RLC}}

The gyrator is a passive, linear, lossless, nonreciprocal, two-port
electrical element. It allows network realizations of devices that
cannot be realized with the conventional four components (i.e., resistors,
inductors, capacitors, and transformers) \cite{tellegen1948gyrator,shenoi1965practical}.
An important property of a gyrator is that it inverts the current-voltage
characteristic; therefore, an impedance load is also inverted across
the gyrator. In other words, a gyrator can make a capacitive circuit
behave inductively, and a series LC circuit behaves like a parallel
LC circuit. The instantaneous voltages and currents on the gyrator
ports are related by \cite{tellegen1948gyrator}

\begin{figure}[!h]
\centering{}\includegraphics[width=1\columnwidth]{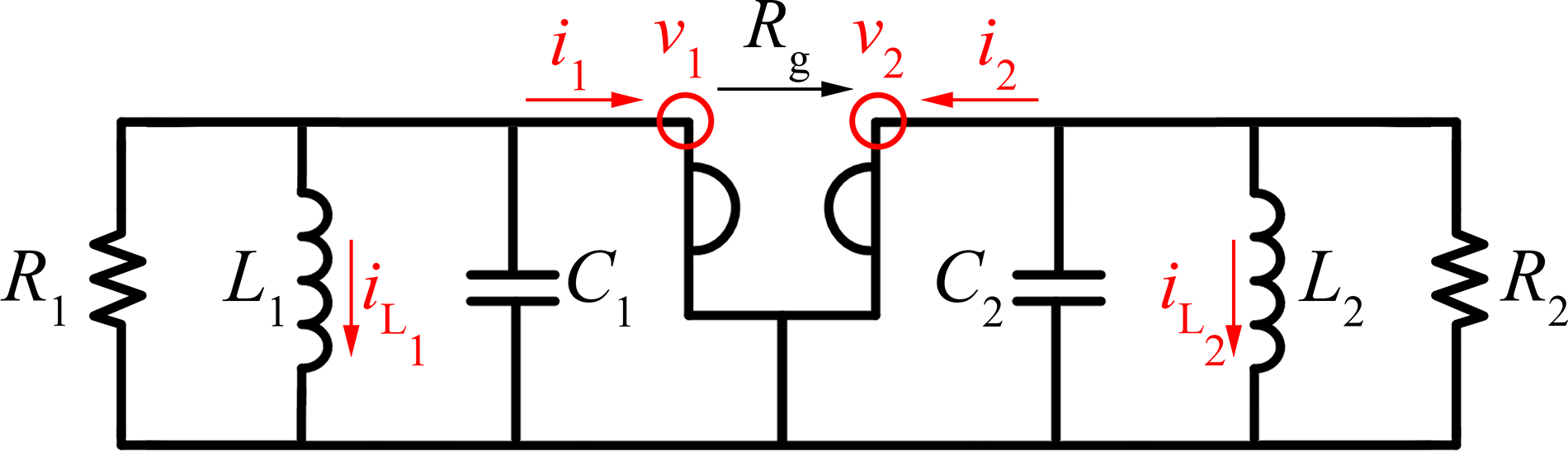}\caption{Schematic view of the lossy parallel-parallel configuration including
losses in each resonator. Inductance and capacitance are negative
in the right resonator.\label{Fig: LossyPPCircuit}}
\end{figure}

\begin{equation}
\left\{ \begin{array}{l}
v_{2}=R_{\mathrm{g}}i_{1}\\
v_{1}=-R_{\mathrm{g}}i_{2}
\end{array}\right.\label{eq:gyrator}
\end{equation}
where the gyration resistance $R_{\mathrm{g}}$ is the important parameter
in the ideal gyrator. In the parallel-parallel configuration, two
parallel RLC resonators are coupled by a gyrator as displayed in Fig.
\ref{Fig: LossyPPCircuit}. We find the EPD condition in this circuit
by writing the Kirchhoff current law equations and finding the associated
Liouvillian matrix. Hence, we assume that all components are ideal,
and inductors and capacitors contain no additional resistance. We
write the two Kirchhoff current law equations and by using the state
vector as, $\boldsymbol{\Psi}\equiv\left[Q_{1},Q_{2},\dot{Q}_{1},\dot{Q}_{2}\right]^{\mathrm{T}}$,
where $Q_{n}$ is the stored charge in the capacitor $C_{n}$ ($n=1$
for the left resonator and $n=2$ for the right resonator), and the
superscript $\mathrm{T}$ denotes the transpose operation. The circuit
dynamics are described based on the Liouvillian formalism as

\begin{equation}
\mathcal{\mathrm{\frac{d\boldsymbol{\Psi}}{dt}=\mathbf{\underline{M}}\boldsymbol{\Psi},\;\;}\mathbf{\underline{M}}}=\left(\begin{array}{cccc}
0 & 0 & 1 & 0\\
0 & 0 & 0 & 1\\
-\omega_{01}^{2} & 0 & -\gamma_{1} & \frac{1}{R_{\mathrm{g}}C_{2}}\\
0 & -\omega_{02}^{2} & -\frac{1}{R_{\mathrm{g}}C_{1}} & -\gamma_{2}
\end{array}\right),
\end{equation}
where $\underline{\mathbf{M}}$ is the $4\times4$ circuit matrix,
and $\gamma_{1}=1/\left(R_{1}C_{1}\right)$ and $\gamma_{2}=1/\left(R_{2}C_{2}\right)$
represent the resonators loss parameter (losses on the right resonator
are represented by a negative $\gamma_{2}$ since $C_{2}$ is negative).
Furthermore, $\omega_{01}=1/\sqrt{C_{1}L_{1}}$, and $\omega_{02}=1/\sqrt{C_{2}L_{2}}$
are resonance angular frequencies of two isolated left and right resonators,
assumed to be both real (the case where they are imaginary is shown
in Ref. \cite{rouhi2022exceptional}. Assuming signals of the form
$Q_{n}\varpropto e^{j\omega t}$, we write the associated eigenvalue
problem, and the characteristic equation is obtained from $\det\left(\underline{\mathbf{M}}-j\omega\underline{\mathbf{I}}\right)=0$,
where $\underline{\mathbf{I}}$ is the identity matrix, leading to

\begin{equation}
\begin{array}{c}
\omega^{4}-j\omega^{3}\left(\gamma_{1}-\gamma_{2}\right)-\omega^{2}\left(\omega_{01}^{2}+\omega_{02}^{2}+\gamma_{1}\gamma_{2}+\frac{1}{R_{\mathrm{g}}C_{1}C_{2}}\right)\\
+j\omega\left(\gamma_{1}\omega_{02}^{2}+\gamma_{2}\omega_{01}^{2}\right)+\omega_{01}^{2}\omega_{02}^{2}=0.
\end{array}\label{eq:PPLossyCharEq}
\end{equation}
The coefficients of the odd-power terms of the angular eigenfrequency
($\omega$ and $\omega^{3}$) in the characteristic equation of Eq.
(\ref{eq:PPLossyCharEq}) are imaginary. Eigenfrequencies $\omega$
and $-\omega^{*}$ are both roots. In order to have a stable circuit
with real-valued eigenfrequencies, the odd-power terms of the angular
eigenfrequency $-j\omega^{3}\left(\gamma_{1}-\gamma_{2}\right)$ and
$j\omega\left(\gamma_{1}\omega_{02}^{2}+\gamma_{2}\omega_{01}^{2}\right)$
in the characteristic equation of Eq. (\ref{eq:PPLossyCharEq}) should
be zero. The coefficient of the $\omega^{3}$ term is zero when $\gamma_{1}=\gamma_{2}$.
We recall that $\gamma_{2}$ is negative, so the condition $\gamma_{1}=\gamma_{2}$
happens either in absence of losses or when one resonator has gain.
However, under this latter gain condition enabling $\gamma_{1}=\gamma_{2}$,
the coefficient of the $\omega$ term $\gamma_{1}\left(\omega_{02}^{2}+\omega_{01}^{2}\right)$
is non-zero because $\omega_{01}^{2}$ and $\omega_{02}^{2}$ are
both positive, and also in this case it would not be possible to have
purely real eigenfreqiencies. On the other hand, the coefficient of
the $\omega$ term vanishes when $\gamma_{1}/\gamma_{2}=-\omega_{01}^{2}/\omega_{02}^{2}$,
and under this condition, the coefficient of the $\omega^{3}$ term
$\gamma_{1}\left(1+\omega_{02}^{2}/\omega_{01}^{2}\right)$ cannot
vanish. In summary, it is not possible to have all real-valued coefficients
in the characteristic polynomials, unless $\gamma_{1}=\gamma_{2}=0$,
which corresponds to a lossless circuit. In other words, under any
amount of small loss, there is no condition to make both $\omega$
and $\omega^{3}$ coefficients equal to zero, hence the eigenfrequencies
are complex, leading to instabilities that cause oscillations. In
the following subsection, we analyze the eigenfrequency in a lossless
structure to further understand the stability of the lossless structure.

\subsection{EPD in Lossless Parallel Circuit\label{subsec:PP}}

To meet the EPD condition for real valued eigenfrequency, we assume
$\gamma_{1}=\gamma_{2}=0$. Accordingly, the circuit consists of two
lossless parallel LC resonators coupled by a gyrator. The eigenfrequencies
for this case are found by solving 
\begin{equation}
\omega^{4}-\omega^{2}\left(\omega_{01}^{2}+\omega_{02}^{2}+\frac{1}{C_{1}C_{2}R_{\mathrm{g}}}\right)+\omega_{01}^{2}\omega_{02}^{2}=0.\label{eq:PP_Char_Eq}
\end{equation}
All the $\omega$'s coefficients are real hence $\omega$ and $\omega^{*}$
are both roots of the characteristic equation. Moreover, it is a quadratic
equation in $\omega^{2}$, therefore $\omega$ and $-\omega$ are
both solutions. The system's angular eigenfrequencies are

\begin{equation}
\omega_{1,3}=\pm\sqrt{a+b},\;\omega_{2,4}=\pm\sqrt{a-b},\label{eq:PPEigenfrequencies}
\end{equation}

\begin{equation}
a=\frac{1}{2}\left(\omega_{01}^{2}+\omega_{02}^{2}+\frac{1}{C_{1}C_{2}R_{\mathrm{g}}}\right),\label{eq: PPEPD_a}
\end{equation}

\begin{equation}
\begin{array}{c}
b^{2}=a^{2}-\omega_{01}^{2}\omega_{02}^{2}.\end{array}\label{eq: PPEPD_b}
\end{equation}
The EPD is obtained when the resonance frequencies of the circuit
coalesce, i.e., when

\begin{equation}
b=0,\label{eq: PP_EPDCondition}
\end{equation}
which happens when $a^{2}=\omega_{01}^{2}\omega_{02}^{2}$. The positive
EPD angular frequency is then given by $\omega_{\mathrm{e}}=\sqrt{a}$,
where we assume $a>0$. The condition to obtain real value for EPD
frequency is rewritten as

\begin{equation}
\omega_{01}^{2}+\omega_{02}^{2}-\omega_{\mathrm{gp}}^{2}>0,\label{eq:PP_EPD_Cond_real}
\end{equation}
where it has been convenient to define $\omega_{\mathrm{gp}}^{2}=-1/\left(C_{1}C_{2}R_{\mathrm{g}}\right)$
for the parallel-parallel configuration (note that $\omega_{\mathrm{gp}}^{2}>0$
because one capacitor is negative). When both Eq. (\ref{eq: PP_EPDCondition})
and inequality in (\ref{eq:PP_EPD_Cond_real}) are satisfied, two
eigenfrequencies coalesce at a real EPD angular frequency,

\begin{equation}
\begin{array}{c}
\omega_{\mathrm{e}}=\sqrt{\frac{1}{2}\left(\omega_{01}^{2}+\omega_{02}^{2}-\omega_{\mathrm{gp}}^{2}\right)}=\sqrt{\omega_{01}\omega_{02}}.\end{array}\label{eq: PP_EPDFrequency}
\end{equation}

\subsection{Dispersion Relation of Lossless and Lossy Parallel-parallel Configurations\label{subsec:Dispersions relation}}

\begin{figure}[t]
\centering{}\includegraphics[width=1\columnwidth]{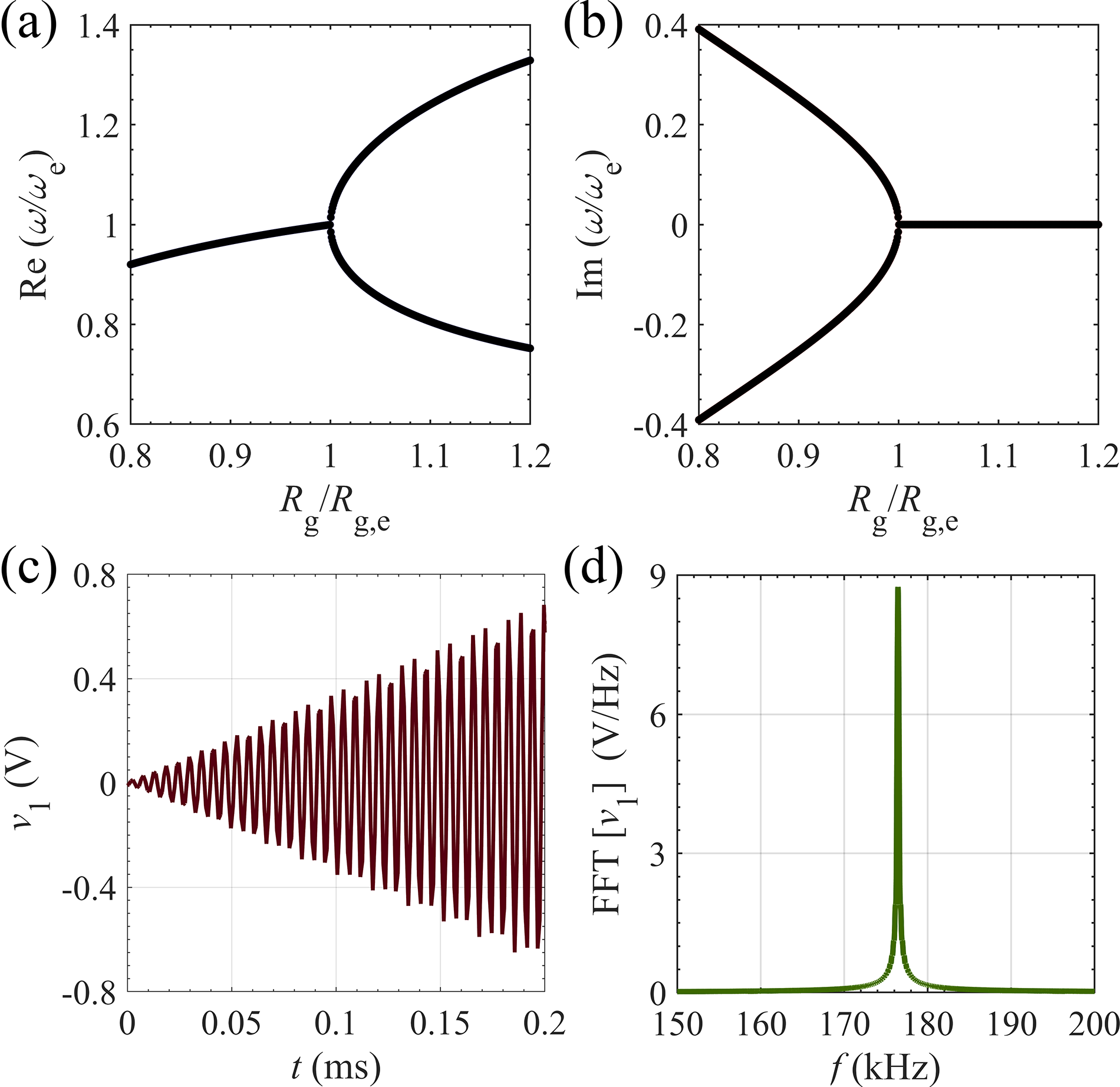}\caption{Variation of the (a) real and (b) imaginary parts of the two eigenfrequencies
to a gyration resistance perturbation in the lossless parallel-parallel
configuration. The bifurcation in the real part is observed for $R_{\mathrm{g}}>R_{\mathrm{g,e}}$.
Voltage $v_{1}$ under the EPD condition in the (c) time domain, and
(d) frequency domain. The frequency domain result is calculated from
$150\mathrm{\:kHz}$ to $200\:\mathrm{kHz}$ performing an FFT with
$10^{6}$ samples in the time window between $0\:\mathrm{ms}$ to
$0.2\:\mathrm{ms}$.\label{Fig: LosslessPPDispersion}}
\end{figure}

As an example, we use the following values: $L_{1}=47\mathrm{\:\mu H}$,
$L_{2}=-47\mathrm{\:\mu H}$, $C_{2}=-47\mathrm{\:nF}$, and $R_{\mathrm{g}}=50\:\Omega$.
We then obtain two values of capacitance $C_{1,\mathrm{e}}=6.34\:\mathrm{nF}$
and $C_{1,\mathrm{e}}=125.25\:\mathrm{nF}$ by imposing Eq. (\ref{eq: PP_EPDCondition})
to be satisfied. Here, both capacitors lead to $a>0$, enabling the
EPD angular frequency to be real valued. Indeed, the $C_{1,\mathrm{e}}=6.34\:\mathrm{nF}$
leads to $\omega_{\mathrm{e}}=1.11\times10^{6}\:\mathrm{rad/s}$,
whereas $C_{1,\mathrm{e}}=125.25\:\mathrm{nF}$ leads to $\omega_{\mathrm{e}}=5.26\times10^{5}\:\mathrm{rad/s}$.
In the following we use $C_{1,\mathrm{e}}=6.34\:\mathrm{nF}$. The
results in Figs. \ref{Fig: LosslessPPDispersion}(a) and (b) show
the branches of the real and imaginary parts of perturbed eigenfrequencies
obtained from the eigenvalue problem when varying the gyrator resistance
near $R_{\mathrm{g,e}}=50\:\Omega$. The bifurcation of the real part
in this case happens for $R_{\mathrm{g}}>R_{\mathrm{g,e}}$ . Perturbing
other components like $C_{1}$ or $L_{1}$ leads to analogous results.

The time domain simulation result for the node voltage $v_{1}$ in
Fig. \ref{Fig: LosslessPPDispersion}(c) is obtained using the Keysight
ADS circuit simulator by employing the ideal model for the gyrator,
using the above circuit values that lead to the EPD. We assume the
capacitor has an initial voltage on $C_{1}$ equal to $1\:\mathrm{mV}$.
The voltage grows linearly with increasing time, demonstrating two
eigenvalues of the circuit are coalescing, and the system exhibits
a double pole, as shown later on. This is peculiar of a second-order
EPD. The spectrum of the voltage $v_{1}$ in Fig. \ref{Fig: LosslessPPDispersion}(d)
is calculated after performing the FFT. The oscillation frequency
is $f_{\mathrm{o}}=176.66\mathrm{\:kHz}$, which is the EPD frequency
calculated above.

By perturbing the gyration resistance, the circuit no longer operates
at EPD. For a higher gyration resistance value, $R_{\mathrm{g}}=52.5\:\Omega>R_{\mathrm{g,e}}=50\:\Omega$
as a $5\%$ increase, we obtain two distinct real-valued eigenfrequencies
in the system. Thus, we could estimate the amount of perturbation
in $R_{\mathrm{g}}$ by measuring the frequency of these two resonances.
On the other hand, by reducing the amount of perturbed parameter by
$5\%$, leading to $R_{\mathrm{g}}=47.5\:\Omega<R_{\mathrm{g,e}}=50\:\Omega$,
the system has two complex eigenfrequencies with non-zero imaginary
parts. The circuit contains signals that are damping or growing exponentially.

\begin{figure}[t]
\centering{}\includegraphics[width=1\columnwidth]{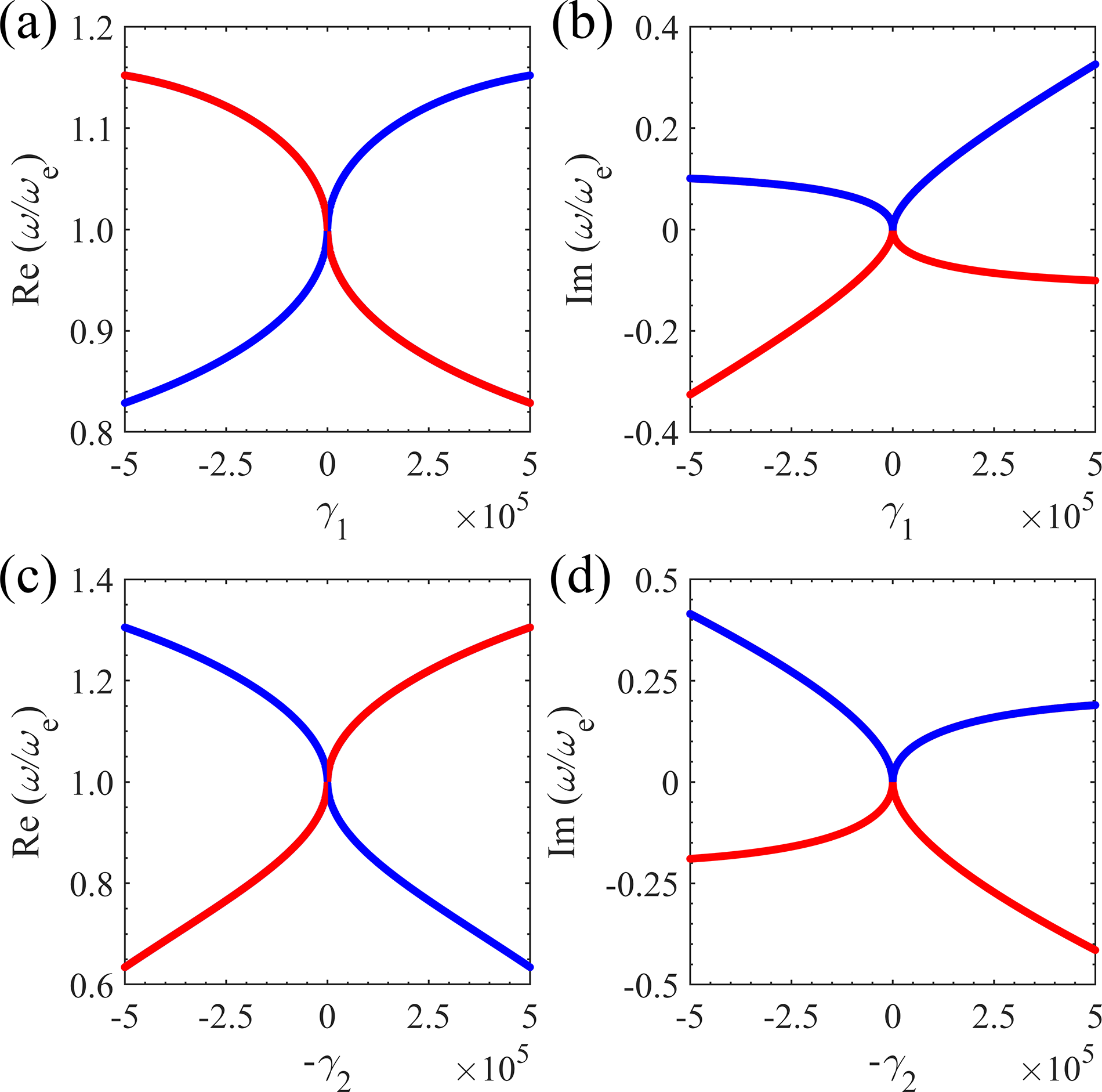}\caption{Variation of (a) real and (b) imaginary parts of the angular eigenfrequencies
to a resistor perturbation on the left resonator. In these plots,
$\gamma_{1}$ is varied whereas we assume $\gamma_{2}=0$. Variation
of (c) real and (d) imaginary parts of the angular eigenfrequencies
to a resistor perturbation on the right resonator. In these plots,
$-\gamma_{2}$ is varied whereas we assume $\gamma_{1}=0$. In these
plots, blue curves show stable branches with positive imaginary parts
and red curves show unstable branches with negative imaginary parts.
In addition, the right half of each plot demonstrates the variation
in eigenfrequencies due to varying positive resistance, whereas the
left half demonstrates the variation in eigenfrequencies due to varying
negative resistance.\label{Fig: LossyPPDispersion}}
\end{figure}

In the lossy circuit, we use the same values of lossless parallel-parallel
configuration for the resonators and gyration resistance plus the
two resistances. In Figs. \ref{Fig: LossyPPDispersion}(a) and (b),
we vary $\gamma_{1}$ and assume $\gamma_{2}=0$, whereas in Figs.
\ref{Fig: LossyPPDispersion}(c) and (d), we perturb $-\gamma_{2}$
and assume $\gamma_{1}=0$. In fact, when we vary $\gamma_{1}$ or
$\gamma_{2}$, we actually vary $R_{1}$ or $R_{2}$, while keeping
constant $C_{1}$ and $C_{2}$. When $\gamma_{1}=\gamma_{2}=0$, the
EPD frequency is the same as the one found earlier for the lossless
configuration in Section \ref{subsec:PP}. Figures \ref{Fig: LossyPPDispersion}(a)-(d)
show the bifurcation of the real and imaginary parts of eigenfrequencies
on both sides of the EPD. It means that the circuit is very sensitive
to both positive and negative variations in the resistance value.
The angular eigenfrequencies are complex-valued for any amount of
loss and the circuit is in the self-oscillation regime. The circuit's
signal oscillates with the frequency associated with the real part
of the unstable eigenfrequency, and the signal grows exponentially
based on the unstable imaginary part of the eigenfrequency. The calculated
results show that we achieve higher sensitivity when perturbing $\gamma_{2}$.

\subsection{Parallel Lossless Circuit Sensitivity}

\begin{figure}[t]
\centering{}\includegraphics[width=1\columnwidth]{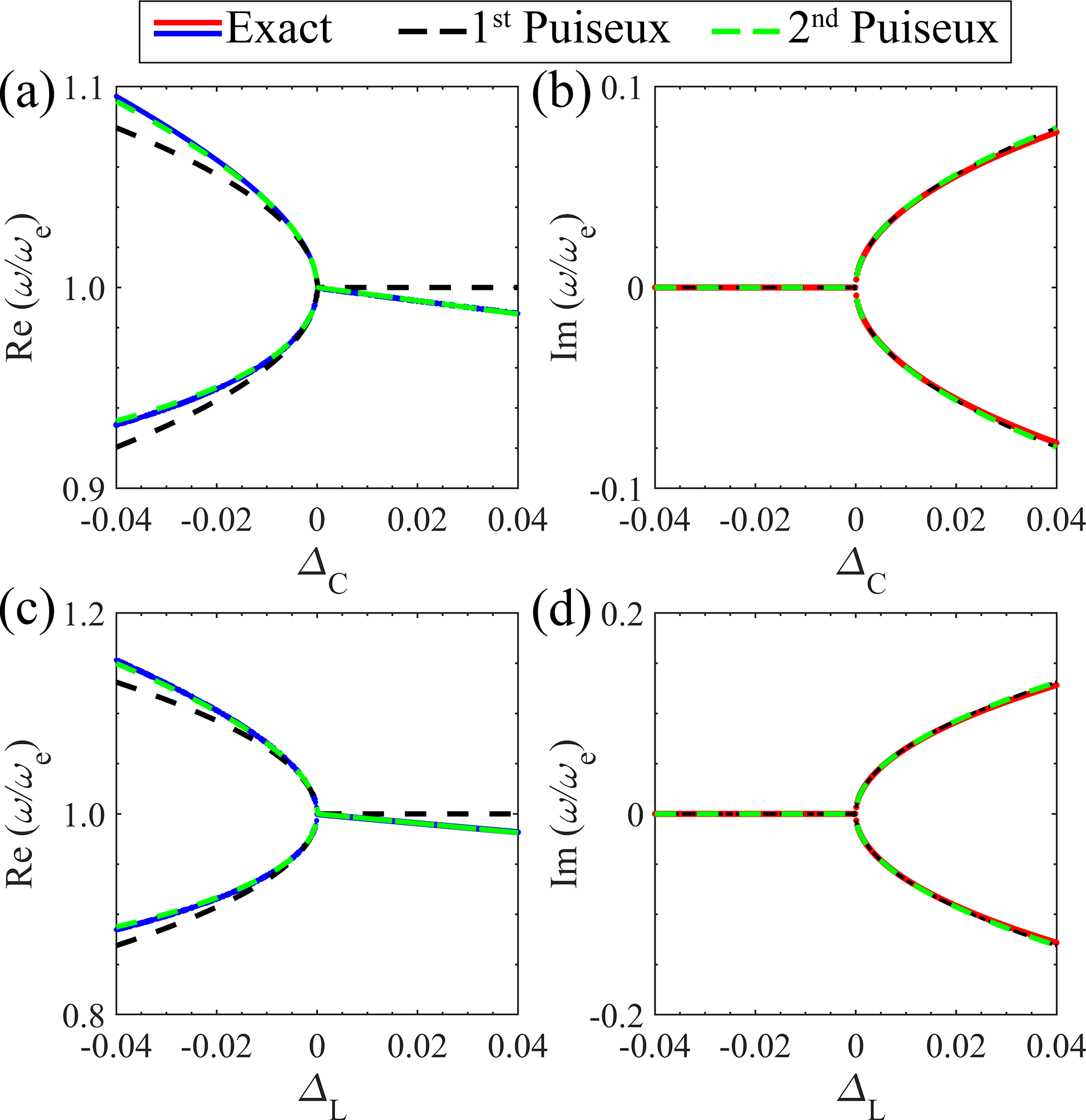}\caption{High sensitivity of the (a) real and (b) imaginary parts of the eigenfrequencies
to relative capacitance perturbation $\Delta_{\mathrm{C}}=(C_{\mathrm{1}}-C_{\mathrm{1,e}})/C_{\mathrm{1,e}}$.
The two perturbed frequencies are real for $\Delta_{\mathrm{C}}<0$.
High sensitivity of the (c) real and (d) imaginary parts of the eigenfrequencies
to relative inductance perturbation $\Delta_{\mathrm{L}}=\left(L_{\mathrm{1}}-L_{\mathrm{1,e}}\right)/L_{\mathrm{1,e}}$.
The two perturbed frequencies are real for $\Delta_{\mathrm{L}}<0$.\label{Fig: Sensitivity}}
\end{figure}

The degenerate eigenvalue (resonance frequency) at an EPD is exceedingly
sensitive to perturbations of system parameters. Here, we show that
the sensitivity of a system\textquoteright s observable to a specific
variation of a component's value is large because of EPD. Let us consider
the parallel-parallel configuration in the EPD regime, with the values
of the components given in Section \ref{subsec:PP}. We select the
parallel case because all elements are grounded and this sometimes
represents a simplification when using realistic active components
that require biasing (for more information on dispersion relation
for series-series configuration you can refer to the Appendix \ref{subsec: Circuits-duality}).
For simplicity, we discuss the case without resistances and we define
the relative circuit perturbation $\Delta_{\mathrm{X}}$ as

\begin{equation}
\Delta_{\mathrm{X}}=\frac{X-X_{\mathrm{e}}}{X_{\mathrm{e}}},\label{eq: perturbation_touchstone}
\end{equation}
where $X$ is the perturbed value of a component and $X_{\mathrm{e}}$
is the unperturbed value that provides the EPD. The subscript ``X''
denotes the perturbed parameter. In this section, we consider variations
of $C_{\mathrm{1}}$ and $L_{\mathrm{1}}$, one at a time, in the
lossless configuration. The calculated diagrams for the real and imaginary
parts of the eigenfrequencies near the EPD are shown in Fig. \ref{Fig: Sensitivity}.
We conclude that the individual variation of the parameters of $C_{\mathrm{1}}$
or $L_{\mathrm{1}}$ show similar sensitivity behavior, i.e., the
real part of the eigenfrequencies splits for $\Delta_{\mathrm{X}}<0$.
Note that the $L_{\mathrm{1}}$ perturbation shows higher sensitivity
because of the wider bifurcation in the dispersion diagram.

We explain the extreme sensitivity by resorting to the general theory
of EPDs. Note that after applying a perturbation in $\Delta_{\mathrm{X}}$
value, we will have a perturbed matrix $\mathbf{\underline{M}}\left(\Delta_{\mathrm{X}}\right)$.
Consequently, the two degenerate eigenvalues at the EPD change considerably
due to the small perturbation in $\Delta_{\mathrm{X}}$, resulting
in two distinct eigenfrequencies $\omega_{\mathrm{\mathit{p}}}\left(\Delta_{\mathrm{X}}\right)$,
with $p=1,2$. A single convergent Puiseux series is used to represent
the two perturbed eigenvalues near an EPD, where the coefficients
are calculated using the explicit recursive formulas presented in
\cite{Welters2011Explicit}. An approximation of $\omega_{p}\left(\Delta_{\mathrm{X}}\right)$
around a second-order EPD is given by

\begin{equation}
\omega_{p}\left(\Delta_{\mathrm{X}}\right)\approx\omega_{\mathrm{e}}+\left(-1\right)^{p}\alpha_{1}\sqrt{\Delta_{\mathrm{X}}}+\alpha_{2}\Delta_{\mathrm{X}}.\label{eq:Puiseux}
\end{equation}
Following \cite{Welters2011Explicit}, we calculate the coefficients
as

\begin{equation}
\alpha_{1}=\sqrt{-\frac{\frac{\partial H\left(\Delta_{\mathrm{X}},\omega\right)}{\partial\Delta_{\mathrm{X}}}}{\frac{1}{2!}\frac{\partial^{2}H\left(\Delta_{\mathrm{X}},\omega\right)}{\partial\omega^{2}}}},\label{eq:PuiseuxCoeff1}
\end{equation}

\begin{equation}
\alpha_{2}=-\frac{\alpha_{1}^{3}\frac{1}{3!}\frac{\partial^{3}H\left(\Delta_{\mathrm{X}},\omega\right)}{\partial\omega^{3}}+\alpha_{1}\frac{\partial^{2}H\left(\Delta_{\mathrm{X}},\omega\right)}{\partial\omega\partial\Delta_{\mathrm{X}}}}{\alpha_{1}\frac{\partial^{2}H\left(\Delta_{\mathrm{X}},\omega\right)}{\partial\omega^{2}}},\label{eq: PuiseuxCoeff2}
\end{equation}
evaluated at the EPD, i.e., at $\Delta_{\mathrm{X}}=0$ and $\omega=\omega_{\mathrm{e}}$,
where $H\left(\Delta_{\mathrm{X}},\omega\right)=\mathrm{det}\left[\mathbf{\underline{M}}\left(\Delta_{\mathrm{X}}\right)-j\omega\underline{\mathbf{I}}\right]$.
Equation (\ref{eq:Puiseux}) indicates that for a small perturbation
$\Delta_{\mathrm{X}}\ll1,$ the eigenvalues change dramatically from
their original degenerate value due to the square root function. In
the first example, the perturbed parameter is the positive capacitance
on the left resonator, $\Delta_{\mathrm{C}}=\left(C_{\mathrm{1}}-C_{\mathrm{1,e}}\right)/C_{\mathrm{1,e}}$,
and the Puiseux series first-order coefficient is calculated by Eq.
(\ref{eq:PuiseuxCoeff1-C}) as $\alpha_{1}=j4.4138\times10^{5}\:\mathrm{rad/s}$
and the second-order coefficient is calculated as $\alpha_{2}=-3.6503\times10^{5}\:\mathrm{rad/s}$.
The result in Figs. \ref{Fig: Sensitivity}(a) and (b) shows the two
branches of the perturbed eigenfrequencies $\omega$ obtained directly
from the eigenvalue problem (or characteristic equation) when the
perturbation $\Delta_{\mathrm{C}}$ is applied. This example uses
$C_{1}$ as a sensing component to detect variation in physical or
chemical parameters converted into electrical parameters. However,
variation in other components' values can also be used in various
realistic scenarios. Figures \ref{Fig: Sensitivity}(a) and (b) demonstrate
that the perturbed eigenvalues can be estimated with great accuracy
by using the Puiseux series truncated at its second order (green dashed
lines). We have also shown the first-order approximation in Figs.
\ref{Fig: Sensitivity}(a) and (b) for better comparison (black dashed
lines), which is also in good agreement with the eigenfrequencies
obtained by the characteristic equation. For a small positive value
of $\Delta_{\mathrm{C}}$, the imaginary parts of the eigenfrequencies
experience a sharp change, while their real parts remain more or less
constant. A small negative value of $\Delta_{\mathrm{C}}$ causes
a rapid variation in the real part of the eigenfrequencies.

In the second example, the perturbed parameter is the positive inductance
on the left resonator, $\Delta_{\mathrm{L}}=\left(L_{\mathrm{1}}-L_{\mathrm{1,e}}\right)/L_{\mathrm{1,e}}$,
and the Puiseux series first-order coefficient is calculated by Eq.
(\ref{eq:PuiseuxCoeff1-L}) as $\alpha_{1}=j7.2792\times10^{5}\:\mathrm{rad/s}$
and the second-order coefficient is calculated as $\alpha_{2}=-5.1598\times10^{5}\:\mathrm{rad/s}$.
The calculated results in Fig. \ref{Fig: Sensitivity}(c) and (d)
show the two branches of the perturbed eigenfrequencies obtained from
the eigenvalue problem when the perturbation in inductance is applied.
By applying the Puiseux series truncated at its second order, it is
possible to estimate the perturbed eigenfrequencies with high accuracy.
However, the first order also provides relatively accurate results.
The imaginary parts of the eigenfrequencies undergo a sharp change
for very small positive perturbations, while their real parts remain
relatively unchanged. A small negative perturbation in the inductor
value causes rapid variation in the eigenfrequencies' real part. This
feature is one of the most extraordinary physical properties associated
with the EPD and it can be exploited for designing ultra-sensitive
sensors \cite{Hodaei2017Enhanced}.

\subsection{Frequency Domain Analysis of The Degenerate Resonance\label{subsec:Frequency-Domain-Analysis}}

\begin{figure}[t]
\centering{}\includegraphics[width=1\columnwidth]{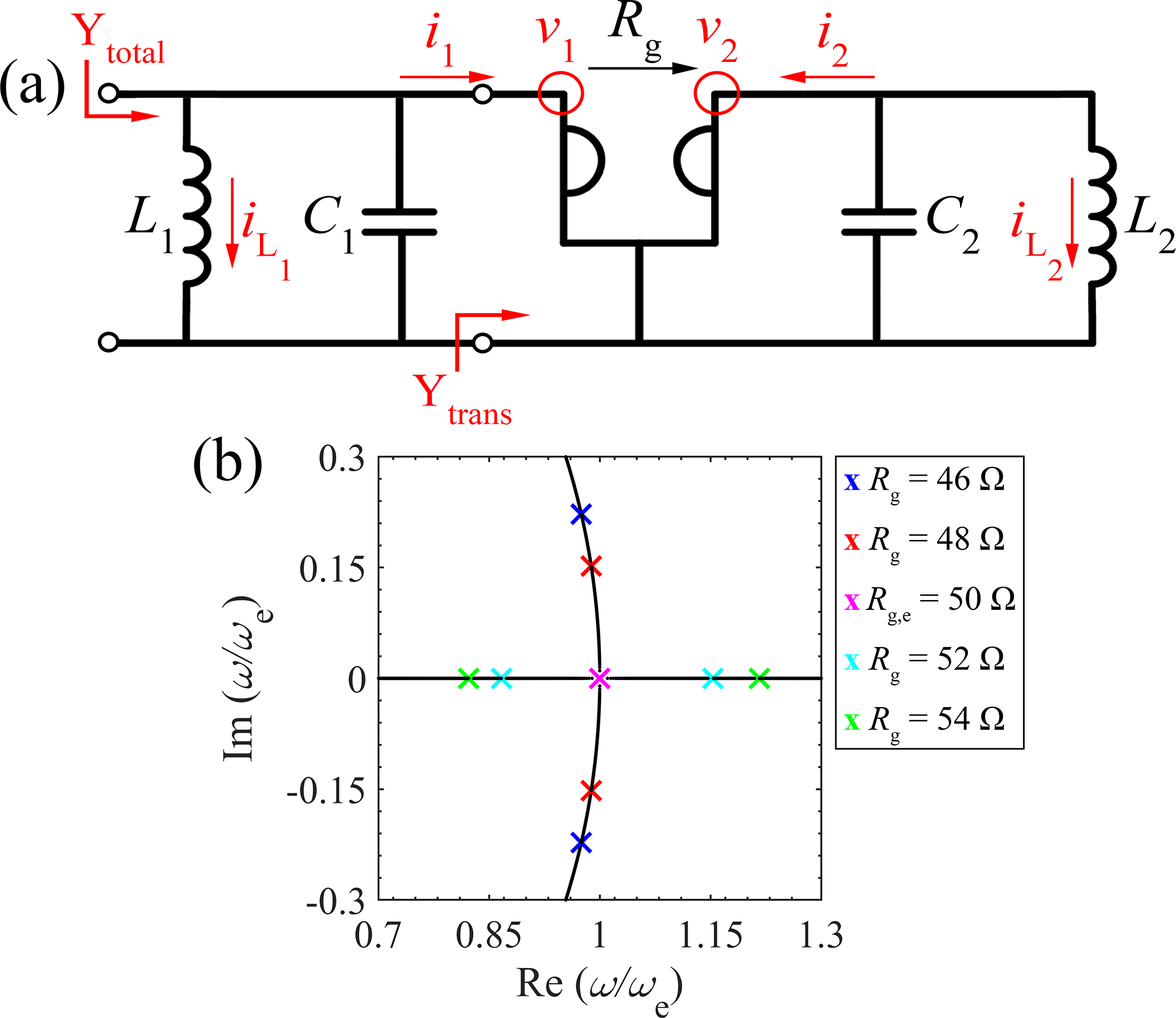}\caption{(a) Schematic view of the lossless parallel-parallel configuration.
(b) Root locus of zeros of $Y_{\mathrm{total}}$ shows the real and
imaginary parts of resonance frequencies of the parallel configuration
when varying gyration resistance. The EPD frequency corresponds to
a double zero of the admittance $Y_{\mathrm{total}}$.\label{Fig: PPRootLucas}}
\end{figure}

We show how the EPD regime is associated to a special kind of circuit's
resonance, directly observed in a frequency domain analysis of the
circuit. We calculate the total input admittance, $Y_{\mathrm{total}}\left(\omega\right)$
(see Fig. \ref{Fig: PPRootLucas}(a)), for the parallel-parallel circuit
by finding the transferred impedance $Y_{\mathrm{trans}}\left(\omega\right)$
on the left side of the circuit. We define the two admittances of
the resonators as $Y_{1}=j\omega C_{1}+1/\left(j\omega L_{1}\right)$,
$Y_{2}=j\omega C_{2}+1/\left(j\omega L_{2}\right)$, and calculate
the transferred admittance on the left side as

\begin{equation}
Y_{\mathrm{trans}}\left(\omega\right)=\frac{1}{R_{\mathrm{g}}^{2}Y_{2}}.
\end{equation}
The total admittance $Y_{\mathrm{total}}\left(\omega\right)$ is calculated
as

\begin{equation}
Y_{\mathrm{total}}\left(\omega\right)\triangleq Y_{1}\left(\omega\right)+Y_{\mathrm{trans}}\left(\omega\right)=Y_{1}+\frac{1}{R_{\mathrm{g}}^{2}Y_{2}}.\label{eq: Ytotal}
\end{equation}
The resonant angular frequencies are obtained imposing $Y_{\mathrm{total}}\left(\omega\right)=0$.
A few steps lead to the same $\omega$-zeros given by Eq. (\ref{eq:PPEigenfrequencies}).
We calculate the resonance frequencies for various gyration resistance
values in Fig. \ref{Fig: PPRootLucas}(b). When considering the EPD
gyrator resistance $R_{\mathrm{g}}=R_{\mathrm{g,e}}=50\,\Omega$,
one has $Y_{\mathrm{total}}\left(\omega\right)\propto\left(\omega-\omega_{\mathrm{e}}\right)^{2}$,
i.e., the two zeros coincide, represented by the point where the two
curves meet exactly at EPD angular frequency. For $R_{\mathrm{g}}<R_{\mathrm{g,e}}$,
resonance angular frequencies are complex conjugate pairs and for
$R_{\mathrm{g}}>R_{\mathrm{g,e}}$, the resonance angular frequencies
are purely real, consistent with the results in Fig. \ref{Fig: LosslessPPDispersion}(a)
and (b).

\section{Experimental Sensitivity in the Saturated Regime \label{sec:Measurement}}

We explore experimentally what happens in the proposed second-order
EPD circuit due to unavoidable instabilities, and we actually exploit
them by making an oscillator. The key experimental observations are
that we have stable oscillations after reaching saturation and the
oscillation frequency exhibits high sensitivity to perturbations.
We also show that such sensitivity is higher than that of a perturbed
single LC resonator. We begin with the linear case, detailing the
design of the different components in the circuit. Furthermore, we
test the proposed circuit for each part, such as the gyrator implementation
with op amps. Next, we study the gyrator-based circuit in the saturation
regime, where the oscillation occurs due to the nonlinearity induced
by op amps and losses/gains in each resonator. Finally, we analyze
the circuit's sensitivity to capacitance changes. The values for the
experiment are set as $C_{1}=400\:\mathrm{nF}$, $L_{1}=47\:\mathrm{\mu H}$,
$L_{2}=-10\:\mathrm{\mu H}$ and $C_{2}=-470\:\mathrm{nF}$ where
these values ideally lead to an EPD at $f=51.9\:\mathrm{kHz}$.

\subsection{Observation of Instability in The Circuit\label{subsec:measurment_linear}}

\begin{figure}[t]
\begin{centering}
\includegraphics[width=1\columnwidth]{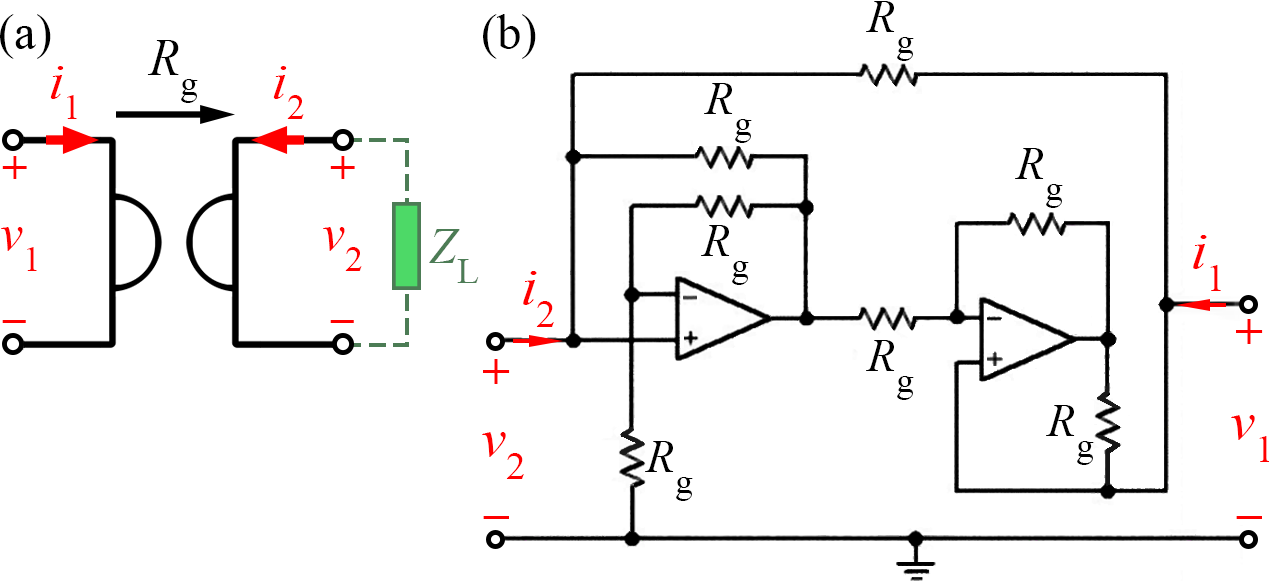}
\par\end{centering}
\caption{(a) Gyrator schematic and corresponding voltages, currents, and gyration
resistance direction. (b) A possible circuit for a gyrator implementation
by using two op amps and seven resistors.\label{Fig: Gyrator}}
\end{figure}

The gyration resistance has an associated direction indicated by an
arrow in the schematic illustrated in Fig. \ref{Fig: Gyrator}(a).
Consider the gyrator circuit shown in Fig. \ref{Fig: Gyrator}(b)
implemented using two op amps of the same model (Texas Instruments,
model TLE2071ACP) and seven resistors. By selecting the proper value
for the resistors (i.e., $R_{\mathrm{g}}$ in Fig. \ref{Fig: Gyrator}(b)),
we control the gyration parameter $R_{\mathrm{g}}$. The gyrator asymmetric
impedance matrix $\mathbf{\underline{Z}}_{\mathrm{g}}$ is given by
Eq. (\ref{eq:gyrator}). To achieve a gyration $R_{\mathrm{g}}=10\:\Omega$,
all the resistances in the circuit shown in Fig. \ref{Fig: Gyrator}(b)
are set to the same value of $10\:\Omega$. We tested the gyrator
circuit shown in Fig. \ref{Fig: Gyrator}(b) with different loads
to ensure that it works properly. We put a load of $Z_{\mathrm{L}}=33\:\Omega$
on the right port of gyrator and measure the transferred impedance
on the left port as $Z_{\mathrm{trans}}=3.56\:\Omega$ and $Z_{\mathrm{trans}}=3.42\:\Omega$,
at frequency of $10$ and $100$ kHz, respectively, showing an experimental
gyration of approximately $9.3\:\Omega$ and $9.65\:\Omega$, respectively.
The impedances are measured with an LCR meter (Keysight, model U1733C).

To satisfy the EPD condition based on the theory discussed in Subsection
\ref{subsec:PP}, the positive capacitance in the experiment $C_{1}$
is built by paralleling fixed capacitors with values $220\:\mathrm{nF}$,
$150\:\mathrm{nF}$, $22\:\mathrm{nF}$ and a trimmer (FTVOGUE , model
Variable Capacitance Kit) to reach the value of $400\:\mathrm{nF}$.
The inductance $L_{1}$ is built by using a commercial inductor of
$47\:\mathrm{\mu H}$ (Coilcraft, model MSS7348-473) with a series
DC loss of $0.15\:\Omega$ and $\pm20\%$ tolerances. However, due
to tolerances, our chosen inductor had a smaller value than the nominal
one. We then tuned the inductance by a series inductor with a value
of $1\:\mathrm{\mu H}$ (Bourns, model 78F1R0K-TR-RC), reaching the
measured value of $46.8\:\mathrm{\mu H}$ at $100$ kHz.

The negative capacitance and inductance are implemented based on the
circuits shown in Appendix \ref{subsec:The-impedance-inverter} using
the same op amp model of the gyrator. In particular, the negative
inductance $L_{2}$ is built by using a commercial inductor of $10\:\mathrm{\mu H}$
(Coilcraft, model MSS7348-103MEC) with a series DC loss $0.045\:\Omega$
and $\pm20\%$ tolerances followed by an inverter in Appendix \ref{subsec:The-impedance-inverter}.
However, due to tolerances, our chosen inductor had a smaller value
than the nominal one. Then, we tuned the inductance by adding a series
inductor with a value of $1\:\mathrm{\mu H}$ (Bourns, model 78F1R0K-TR-RC)
and a $0.47\:\mathrm{\mu H}$ (Bourns, model 542-78FR47K-RC) reaching
a measured value of $10.1\:\mathrm{\mu H}$ at $100$ kHz.

The value of $C_{2}=-470\:\mathrm{nF}$ comes from the inverter described
in Appendix \ref{subsec:The-impedance-inverter}, where we ignored
the gain associated to inversion because of the capacitance's high
quality factor. We use a capacitance trimmer (FTVOGUE , model Variable
Capacitance Kit) in parallel to a commercial fixed capacitor with
a value of $470\:\mathrm{nF}$ to achieve a value close to the desired
capacitance of $470\:\mathrm{nF}$. All capacitances and inductances
are measured with an LCR meter (Keysight, model U1733C) at 100 kHz
to ensure the tuning process leads to the desired capacitances and
inductances design values. The dispersion diagram of the complex eigenfrequencies
by varying the capacitance $C_{1}$ is shown in Fig. \ref{Fig:PPEPDMeasurement-2}.
The system is unstable for any shown value of the perturbed capacitor
because of the non-zero imaginary part of eigenfrequencies.

Due to the presence of loss in the prototype, which is caused mainly
by the two inductors, we do not have the perfect degeneration of the
eigenfrequencies, as shown in Fig. \ref{Fig:PPEPDMeasurement-2}(b)
where we do not have any more $\mathrm{Im}\left(f\right)=0$ near
the bifurcation (plot obtained via simulation in the linear regime).
In addition, in the nonlinear regime, the negative inductance and
capacitance are responsible of the saturation regime because they
act as gain when transferred by the negative impedance converter shown
in Appendix \ref{subsec:The-impedance-inverter}. Nevertheless, we
can still get in the vicinity of the original EPD frequency.

Since the stored charge in the capacitors is $Q_{n}\varpropto e^{j\omega t}$,
even a small negative imaginary part of an eigenfrequency leads to
an instability and the establishment of self substained oscillations,
while the charge associated with the other eigenfrequency with positive
imaginary part decays. In conclusion, the small gain generated by
the inverters leads to non-zero $\mathrm{Im}\left(f\right),$ which
induces the system state vectors to grow exponentially and it results
in an unstable system. We use the instability of the system to our
advantage by letting the system oscillate at the EPD frequency, and
we investigate the sensitivity of the system in the saturation regime
close to the original EPD frequency.

\subsection{Measurements in The Saturation Regime\label{subsec:measurment_sat}}

\begin{figure}[t]
\centering{}\includegraphics[width=0.5\textwidth]{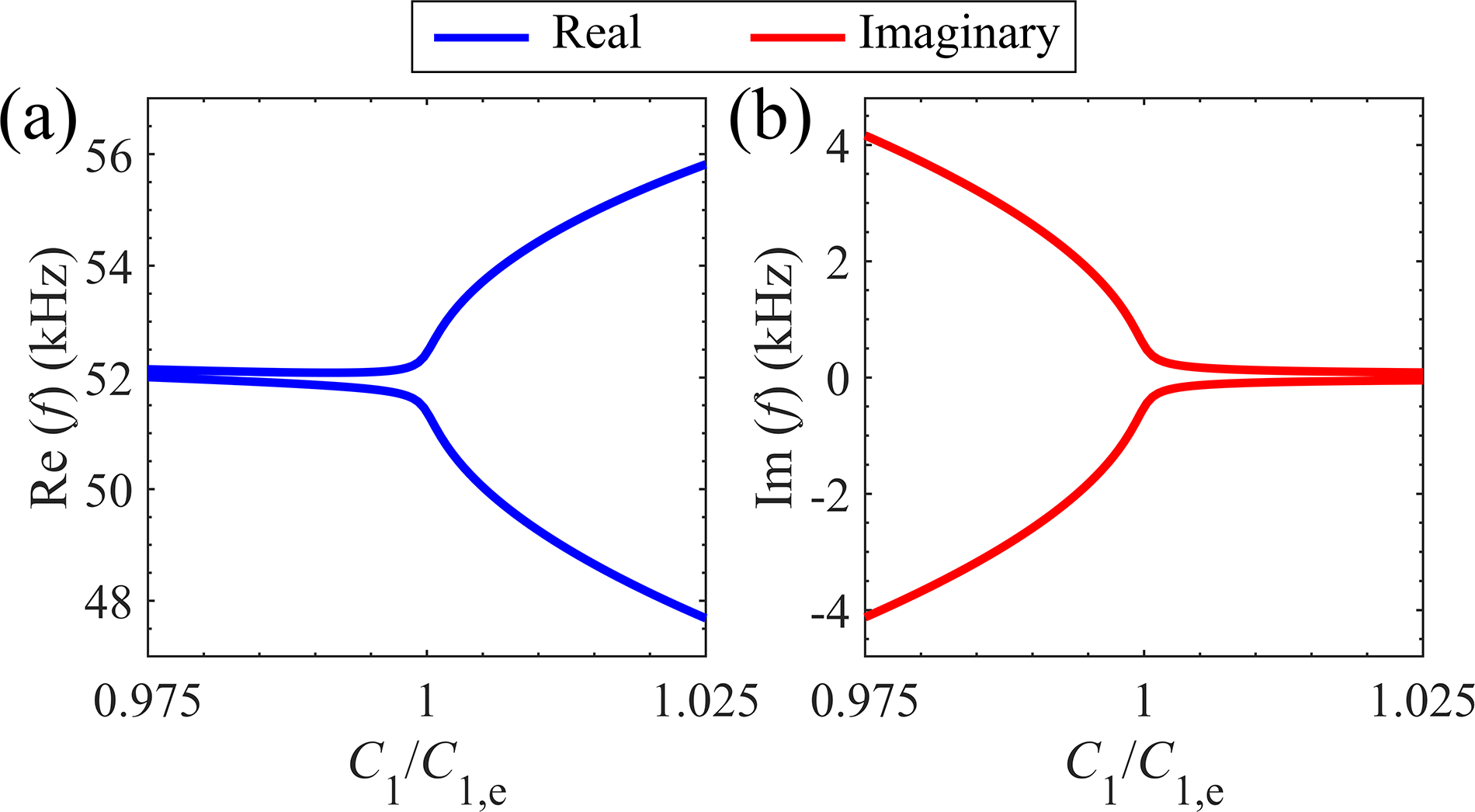}\caption{Variation of (a) real (blue) and (b) imaginary (red) parts of the
eigenfrequencies to a capacitance perturbation on the left resonator.
\label{Fig:PPEPDMeasurement-2}}
\end{figure}

In the experimental setup, the system saturates exhibiting a steady
oscillation at $f_{\mathrm{osc},0}=50.7\:\mathrm{kHz}$, measured
using a spectrum analyzer (Rigol, model DSA832E), which is close to
the frequency shown in Fig. \ref{Fig:PPEPDMeasurement-2} where the
bifurcation is imperfect. We observe that the gyration value is close
to the designed value of $10\:\Omega$ even after reaching saturation
by measuring the voltages $v_{1}$ and $v_{2}$ in the circuit shown
in Fig. \ref{Fig: Gyrator}(a). Also, we considered the transformed
admittance of the positive LCR tank after the gyrator ($1/Y_{\mathrm{trans}}=R_{g}^{2}(C_{1}j\omega\parallel1/(L_{1}j\omega+R_{1}))$
at the frequency of oscillation to find $i_{2}=v_{2}Y_{\mathrm{trans}}$.
As a result, the gyration parameter in the saturated regime at the
oscillation frequency of $f_{\mathrm{osc}}=50.7\:\mathrm{kHz}$ was
calculated as $R_{\mathrm{g}}=10.8\:\Omega$, which is close to the
theoretical design value of $R_{\mathrm{g}}=10\:\Omega$ and from
those measured in the linear regime.

\begin{figure*}[t]
\centering{}\includegraphics[width=0.92\textwidth]{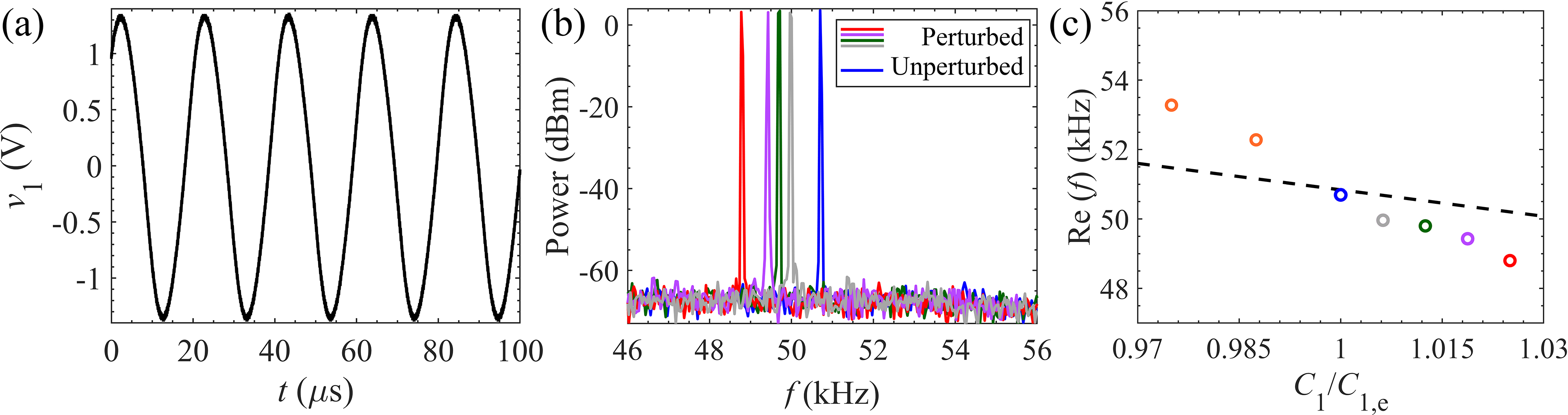}\caption{(a) Measured time-domain voltage signal at the capacitor $C_{1}$
when the system is perturbed from the EPD by $C_{1}-C_{1,0}=2.5\:\mathrm{nF}$.
(b) Measured capacitor voltage spectrum with the unperturbed frequency
of oscillation of $50.7\:\mathrm{kHz}$ (blue); $49.96\:\mathrm{kHz}$
for perturbation $\Delta_{\mathrm{C}}=0.625\:\%$ (gray); $49.8\:\mathrm{kHz}$
for $\Delta_{\mathrm{C}}=1.25\:\%$ (green); $49.43\:\mathrm{kHz}$
for $\Delta_{\mathrm{C}}=1.875\:\%$ (purple); and $48.9\:\mathrm{kHz}$
for $\Delta_{\mathrm{C}}=2.5\:\%$ (red). (c) Comparison of the measured
oscillation frequencies of the proposed nonlinear circuit (color dots)
and those of the perturbed single LC linear circuit (dashed black
line), when perturbing the positive capacitance.\label{Fig:PPEPDMeasurement}}
\end{figure*}

The goal here is to measure the sensitivity of the self oscillation
frequency (after reaching saturation) to the perturbation of the capacitance
$C_{1}$. Indeed, the measured oscillation frequency dramatically
shifts away significantly from the unperturbed frequency $f_{\mathrm{osc},0}=50.7\:\mathrm{kHz}$
even when small perturbations are applied. Figure \ref{Fig:PPEPDMeasurement}(a)
shows the experimental time-domain voltage signal $v_{1}$ of the
capacitor $C_{1}$ with respect to the ground, when a relative perturbation
$\Delta_{\mathrm{C}}=2.5\:\%$ ($\Delta_{\mathrm{C}}=\Delta C/C_{1,0}$,
where subscript $0$ refers to the unperturbed value of $C_{1,0}=400\:\mathrm{nF}$)
is applied to $C_{1}$, measured by an oscilloscope. The voltage frequency
is measured with a spectrum analyzer (Rigol DSA832E), and shown in
Fig. \ref{Fig:PPEPDMeasurement}(b), for various $C_{1}$ values.
A fundamental frequency of oscillation of $f_{\mathrm{osc},0}=50.7\:\mathrm{kHz}$
was observed for no perturbation (blue); $49.96\:\mathrm{kHz}$ for
perturbation $\Delta_{\mathrm{C}}=0.625\:\%$ (gray); $49.8\:\mathrm{kHz}$
for perturbation $\Delta_{\mathrm{C}}=1.25\:\%$ (green); $49.43\:\mathrm{kHz}$
for perturbation $\Delta_{\mathrm{C}}=1.875\:\%$ (purple); and $48.9\:\mathrm{kHz}$
for perturbation $\Delta_{\mathrm{C}}=2.5\:\%$ (red). We used a resolution
bandwidth of 10 Hz, and a video bandwidth of 10 Hz. In all cases,
the measured spectrum is clean approximately down $-70$ dB from the
peak values. The linewidths, calculated at $-3$ dB from the peak,
are approximately $10$ Hz which are significantly smaller than the
measured frequency shifts. The measured oscillation frequencies versus
$C_{1}$ are captured in Fig. \ref{Fig:PPEPDMeasurement}(c) with
circles, with the corresponding colors used in Fig. \ref{Fig:PPEPDMeasurement}(b).
In the experiment, a relative perturbation $\Delta_{\mathrm{C}}=2.5\:\%$
applied to $C_{1}$ in the gyrator-based circuit led to a frequency
shift $\left|\Delta f\right|=\left|48.9\:\mathrm{kHz}-50.7\:\mathrm{kHz}\right|=1.8\:\mathrm{kHz}$,
where $f_{\mathrm{osc},0}=50.7\:\mathrm{kHz}$ is the oscillation
frequency with no perturbation and $f_{\mathrm{osc}}=48.9\:\mathrm{kHz}$
is perturbed oscillation frequency. The measured $-3$-dB (half power)
spectral linewidth of $10\:\mathrm{Hz}$ (using a resolution bandwidth
of 10 Hz, and a video bandwidth of 10 Hz) is 180 times narrower than
the measured frequency shift $\left|\Delta f\right|=1.8\:\mathrm{kHz}$.
For the smallest perturbation of $\Delta_{\mathrm{C}}=0.625\:\%$,
the associated frequency shift $\left|\Delta f\right|=0.74\:\mathrm{kHz}$
is 74 times larger than the linewidth of 10 Hz, indicating that the
perturbed frequency spectrum is clearly detectable (consider also
that the noise floor is 70 dB lower than the peak).

\subsection{Discussion on Sensitivity and Capability to Detect Small Perturbation\label{subsec:DiscussionSensitivity}}

We now elaborate on the sensitivity of the gyrator-based oscillator
in the saturation regime to circuit perturbations and discuss how
the obtained experimental results are comparable to those of (i) a
single LC resonator in the linear regime, and (ii) other circuits
based on EPDs from the literature \cite{Schindler2011Experimental,kazemi2022experimental,Nikzamir2022Highly,Moncada2024Frequency-Domain}.
In particular, we compare the sensitivity $S=\left|\Delta_{\mathrm{f}}\right|/\left|\Delta_{\mathrm{X}}\right|$
in the various cases, where $\Delta_{\mathrm{f}}=\left|\Delta f\right|/f_{0}$
and $\Delta_{\mathrm{X}}=\left|\Delta X\right|/X_{0}$ are the normalized
changes in the perturbed frequency and a generic parameter $X$ in
the circuit (subscript 0 refers to the unperturbed value). The change
in the generic parameter $X$ represents changes in capacitance or
resistance, as will be discussed next.

We first look at the comparison with a single LC resonator in linear
regime, shown in Fig. \ref{Fig:PPEPDMeasurement}(c), where the variation
of the experimental oscillation frequency by perturbing the positive
capacitor $C_{1}$ (colored circles) of the gyrator-based circuit
is compared to the resonant frequency shift of a\textit{ linear single}
LC resonator (dashed black line). The single LC resonator has an inductance
$L_{\mathrm{s}}=24.5\:\mathrm{\mu H}$ and capacitance $C_{\mathrm{s}}=C_{1}=400\:\mathrm{nF}$
(same as $C_{1,0}$ value at EPD), with resonance frequency $f_{0}=1/\left(2\pi\sqrt{L_{\mathrm{s}}C_{\mathrm{s}}}\right)=50.8\:\mathrm{kHz}$.
Perturbing the capacitance $C_{\mathrm{s}}$ leads to a changed frequency
of approximately $f\approx f_{0}\left(1-\Delta_{\mathrm{C}_{\mathrm{s}}}/2\right)$,
where $\Delta_{\mathrm{C}_{\mathrm{s}}}=\left(C-C_{\mathrm{s}}\right)/C_{\mathrm{s}}$.
The results clearly showcase that the gyrator-based circuit in the
saturation regime exhibits higher sensitivity compared to that of
the linear single LC resonator. Notably, the resonance frequency variation
due to the perturbation of capacitance of $\Delta_{\mathrm{C}_{\mathrm{s}}}=10\%$
for the single LC resonator is comparable of the oscillation frequency
shift of only $1/8$ of the perturbation ($\Delta_{\mathrm{C}}=1.25\:\%$)
when using the saturation regime of the gyrator-based oscillator.
This result shows that the sensitivity of our proposed gyrator-based
oscillator is 8 times larger than the one of a single LC resonator.

We now compare the sensitivity of the presented oscillator with those
obtained using the four circuits in Refs. \cite{Schindler2011Experimental,kazemi2022experimental,Nikzamir2022Highly,Moncada2024Frequency-Domain}.
In our developed gyrator-based circuit, we measured a first perturbed
frequency $\left|\Delta_{\mathrm{f}}\right|=1.46\:\%$ for the smallest
capacitance perturbation $\Delta_{\mathrm{C}}=0.625\:\%.$ Hence,
the gyrator-based EPD oscillator exhibits a sensitivity of $S=2.34$,
(for $\Delta_{\mathrm{C}}=0.625\:\%.$) which is comparable to or
higher than the counterpart circuits as shown next. In the linear
single LC resonator, the sensitivity shown in Fig. \ref{Fig:PPEPDMeasurement}(c)
is $S=0.5$, hence it is lower than that of the proposed circuit oscillator.

In Ref. \cite{kazemi2022experimental}, they used a linear regime
in a single time-varying resonator, and they measured a sensitivity
of $\left|\Delta_{\mathrm{f}}\right|=0.66\:\%$, calculated based
on the separation of the two perturbed resonance frequencies, for
their smallest perturbation considered $\Delta_{\mathrm{C}}=0.3\:\%$
, as shown in Fig. 6(a) of their paper. Thus, in \cite{kazemi2022experimental},
they achieved an approximate sensitivity of $S=2.2$ for the first
perturbation.

Sensitivity in the PT-symmetric two-coupled resonators in Ref. \cite{Schindler2011Experimental}
was measured approximately as $S=0.75$ in Fig. 2 of that paper with
$\left|\Delta_{\mathrm{f}}\right|=5\:\%$, calculated based on the
separation of the two resonance frequencies after perturbation, for
their smallest applied perturbation of $\Delta_{\gamma}=6.7\:\%$
where in this case $X=\gamma=R^{-1}\sqrt{L/C}$. Therefore, the highest
sensitivity measured in \cite{Schindler2011Experimental} was just
slightly higher than that of a single linear LC resonator. Since experimental
data at the EPD were not available, the sensitivity was calculated
based on the frequency shift corresponding to perturbation of normalized
$\gamma$ approximately from $1.042$ to $0.975$. These two points
correspond to the first perturbations occurring just before and after
the EPD.

In Ref. \cite{Nikzamir2022Highly}, the authors used an oscillator
scheme using a nonlinear saturated regime (analogous to what considered
in this paper) in two-coupled resonators, and measured a frequency
change of $\left|\Delta_{\mathrm{f}}\right|=1.29\:\%$ for their smallest
perturbation of $\Delta_{\mathrm{C}}=1.3\%$. Therefore, in \cite{Nikzamir2022Highly},
the sensitivity was measured to be $S=0.99$.

In \cite{Moncada2024Frequency-Domain}, the EPD concept using saturable
nonlinear gain was exploited to design oscillators based on two coupled
transmission lines as in \cite{Kazemi2022High}. They observe oscillation
frequency changes in response to perturbations in the load resistance,
so in this case we have $X=R$, and $\Delta_{\mathrm{X}}=\Delta_{\mathrm{R}}$.
In their measurement (Fig. 16 in \cite{Moncada2024Frequency-Domain}),
they observed a $|\Delta_{\mathrm{f}}|=20\%$ for a resistance perturbation
of $\Delta_{\mathrm{R}}=9.8\%$, resulting in a sensitivity of $S=2.04$.
Since experimental data at the EPD were not available, the sensitivity
was calculated based on the frequency shift obtained by perturbing
the load resistance from $51$ $\Omega$ to $56$ $\Omega$, relative
to a fundamental frequency of $1$ GHz.

Note that besides aiming at high sensitivity, another very important
parameter is the \textit{capability to detect small perturbations}.
The measured capacitance variation $\Delta_{\mathrm{C}}=0.625\%$
in the gyrator-based circuit in this paper is higher only than the
case in Ref. \cite{kazemi2022experimental} where $\Delta_{\mathrm{C}}=0.3\:\%$.
Hence, the smallest perturbation in the gyrator-based oscillator is
smaller than the one in \cite{Nikzamir2022Highly} where $\Delta_{\mathrm{C}}=1.3\%$,
and it is much smaller than the smallest variation in \cite{Schindler2011Experimental},
$\Delta_{\mathrm{\gamma}}=6.7\:\%$ (note that the low sensitivity
measured in \cite{Schindler2011Experimental} probably depends on
the large variation they considered, since the EPD-based sensitivity
decreases when moving away from the EPD). Furthermore, the perturbation
measured in \cite{Moncada2024Frequency-Domain} was $\Delta_{\mathrm{R}}=9.8\%$.

The capability to detect small variations $\Delta_{\mathrm{X}}$ depends
on both the \textit{sensitivity} and the \textit{linewidth} of the
oscillator spectrum. The linewidth of the experimental oscillation
frequency was measured to be only $10$ Hz. However, such a linewidth
could be even smaller than that measured value because we did not
have the capability to use a narrower intermediate frequency bandwidth
of the spectrum analyzer to test it.

Nonlinear effects such as gain saturation and Kerr nonlinearity in
systems with EPD play an important role in sensitivity and stability
\cite{Konotop2016Nonlinear}. The analysis of the nonlinear modes
of the system can be carried over by resorting to nonlinear dynamics
tools \cite{kuznetsov1998elements}, where the modes of the system
are identified as fixed points of the system equations. At the same
time, the stability of such fixed points (ultimately determining whether
the system will operate in those modes) is obtained from the Lyapunov
exponents associated to the Jacobian matrix of the system equations
linearized around the fixed point. Some dynamics related to EPDs in
nonlinear systems are discussed in \cite{Zhiyenbayev2019Enhanced,Yu2019Neuromorphic},
or more recently in the framework of electronic circuits, in \cite{suntharalingam2023noise}
where a singularity is found in the nonlinear framework. In \cite{suntharalingam2023noise},
they discussed the sensitivity of the system supporting the EPD and
how nonlinearity helps the signal-noise ratio (how much noise contributes
to the system). In this work, we do not focus on describing the system's
nonlinearity but we observe the saturated regime using an implementation
with electronic components. Based on the experimental results shown
in Fig. \ref{Fig:PPEPDMeasurement}(b) , which shows a clear and sharp
spectrum, for each perturbed case, we conclude that our nonlinear
system works effectively in the saturation regime, for each value
of the perturbed capacitor.

\section{Conclusions}

We have shown that two resonators connected via a gyrator support
an EPD when one resonator is made of a negative inductance and a negative
capacitance. We have provided the theoretical conditions for such
EPD to exist at a purely real frequency and verified our theoretical
calculations by using a time domain circuit simulator (Keysight ADS).
We have demonstrated that the eigenfrequencies are exceptionally sensitive
to a perturbation of the system and this may have significant implications
for ultra-sensitive sensing technology and RF sensors. In addition,
we show that the system has two complex eigenfrequencies, one of which
is always associated to the circuit instability. We have fabricated
the circuit and using the saturated regime, we have measured the sensitivity
of the self-oscillation frequency to small capacitance perturbations.
We have measured both the \textit{sensitivity} and the \textit{linewidth}
of the oscillator's spectrum, because these two parameters are important
for detecting small circuit perturbations. The spectrum exhibited
a very narrow linewidth (i.e., 10 Hz), and the measured signal had
a noise floor at $-70$ dB from the spectrum peak, and the circuit's
sensitivity was measured to be comparable or better than cases previously
published. We believe that the demonstrated concept of an oscillator
in the saturated regime that is very sensitive to perturbations could
pave the way for the development of new operation schemes to boost
the performance of highly sensitive sensors.

\section{Appendix}

\subsection{Circuits Duality\label{subsec: Circuits-duality}}

The concept of duality applies to many fundamental physics/engineering
concepts. For instance, this concept has been utilized many times
in electromagnetic and electric circuits. Two circuits are dual if
the mesh equations that describe one of them have the same mathematical
form as the nodal equations that characterize another circuit \cite{balabanian1969electrical}.
We consider the mesh equations in the parallel-parallel configuration
using the Kirchhoff's voltage law. According to the duality theorem,
if we substitute voltage by current, current by voltage, capacitance
by inductance, and inductance by capacitance, we can obtain mesh equations
for series-series configuration. Thus, we present a dual circuit of
the parallel configuration, achieving an EPD by utilizing gyrator-based
circuits with two series-series resonators.

\begin{figure}[t]
\centering{}\includegraphics[width=1\columnwidth]{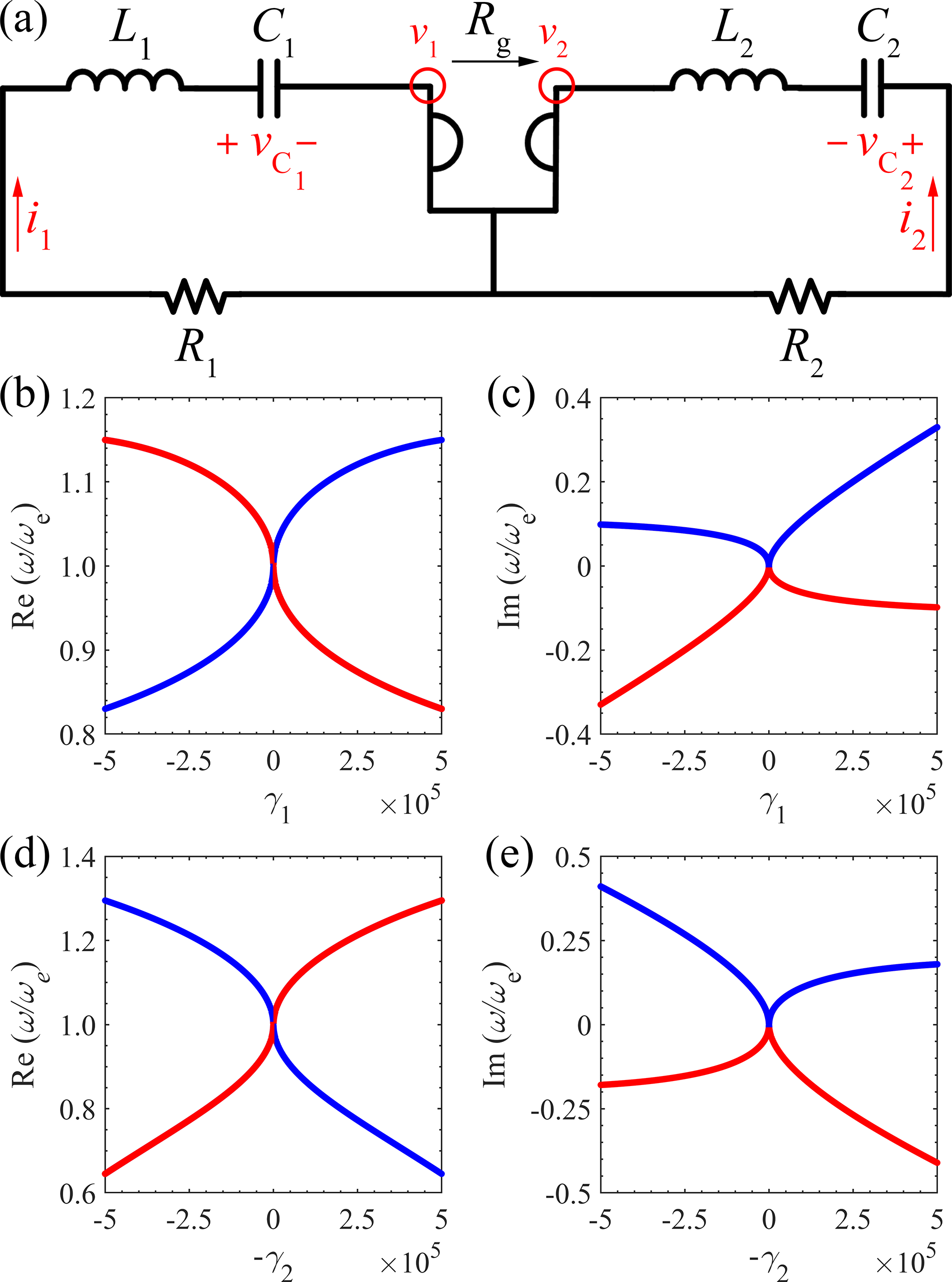}\caption{(a) Schematic view of the lossy series-series configuration including
a resistor in each resonator. The right resonator is made of negative
inductance and capacitance. Variation of (b) real and (c) imaginary
parts of the angular eigenfrequencies to a resistor perturbation in
the left resonator. In these plots, $\gamma_{1}$ is varied whereas
we assume $\gamma_{2}=0$. Variation of (d) real and (e) imaginary
parts of the angular eigenfrequencies to a resistor perturbation on
the right resonator. In these plots, $-\gamma_{2}$ is varied whereas
we assume $\gamma_{1}=0$. In these plots, blue curves show stable
branches with positive imaginary parts and red curves show unstable
branches with negative imaginary parts. The right half of each plot
demonstrates the variation in eigenfrequencies due to varying positive
resistance, whereas the left half demonstrates the variation in eigenfrequencies
due to varying negative resistance.\label{Fig: LossySSCircuitDispersion}}
\end{figure}

\subsubsection{EPD in Lossy Series Circuit\label{subsec:Series-RLC}}

We analyze the EPD condition in the series-series configuration by
accounting for series resistors $R_{1}$ and $R_{2}$ in both resonators.
Using the Liouvillian formalism, the Kirchhoff voltage law equations
for the two loops of the circuit in Fig. \ref{Fig: LossySSCircuitDispersion}(a),
and the state vector of $\boldsymbol{\Psi}\equiv\left[Q_{1},Q_{2},\dot{Q}_{1},\dot{Q}_{2}\right]^{\mathrm{T}}$,
we obtain

\begin{equation}
\mathcal{\mathrm{\frac{d\boldsymbol{\Psi}}{dt}=\mathbf{\underline{M}}\boldsymbol{\Psi},\;\;}\mathbf{\underline{M}}}=\left(\begin{array}{cccc}
0 & 0 & 1 & 0\\
0 & 0 & 0 & 1\\
-\omega_{01}^{2} & 0 & -\gamma_{1} & \frac{R_{\mathrm{g}}}{L_{1}}\\
0 & -\omega_{02}^{2} & -\frac{R_{\mathrm{g}}}{L_{2}} & -\gamma_{2}
\end{array}\right),
\end{equation}
where, $\gamma_{1}=R_{1}/L_{1}$ and $\gamma_{2}=R_{2}/L_{2}$ describe
losses (losses on the right resonator are represented by a negative
$\gamma_{2}$ since $L_{2}$ is negative). These eigenfrequencies
are solutions to the following characteristic equation

\begin{equation}
\begin{array}{c}
\omega^{4}-j\omega^{3}\left(\gamma_{1}-\gamma_{2}\right)-\omega^{2}\left(\omega_{01}^{2}+\omega_{02}^{2}+\gamma_{1}\gamma_{2}+\frac{R_{\mathrm{g}}}{L_{1}L_{2}}\right)\\
+j\omega\left(\gamma_{1}\omega_{02}^{2}+\gamma_{2}\omega_{01}^{2}\right)+\omega_{01}^{2}\omega_{02}^{2}=0.
\end{array}\label{eq:SSLossyCharEq}
\end{equation}

An eigenfrequency with a negative imaginary part is associated with
an exponentially growing signal. The coefficients of the odd-power
terms of the angular eigenfrequency ($\omega$ and $\omega^{3}$)
in the characteristic equation of Eq. (\ref{eq:SSLossyCharEq}) are
imaginary. In the characteristic equation, eigenfrequencies $\omega$
and $-\omega^{*}$ are both roots. In order to have a stable circuit
with real-valued eigenfrequencies the odd-power terms of the angular
eigenfrequency $-j\omega^{3}\left(\gamma_{1}-\gamma_{2}\right)$ and
$j\omega\left(\gamma_{1}\omega_{02}^{2}+\gamma_{2}\omega_{01}^{2}\right)$
in the characteristic equation of Eq. (\ref{eq:SSLossyCharEq}) should
be zero, otherwise a complex eigenfrequency needed to satisfy the
characteristic equation of Eq. (\ref{eq:SSLossyCharEq}). The coefficient
of the $\omega^{3}$ term is zero when $\gamma_{1}=\gamma_{2}$, and
under this condition the coefficient of the $\omega$ term $\gamma_{1}\left(\omega_{02}^{2}+\omega_{01}^{2}\right)$
is non-zero value because $\omega_{01}^{2}$ and $\omega_{02}^{2}$
are both positive. Moreover, the coefficient of the $\omega$ term
vanishes when $\gamma_{1}/\gamma_{2}=-\omega_{01}^{2}/\omega_{02}^{2}$,
and under this condition, the coefficient of the $\omega^{3}$ term
$\gamma_{1}\left(1+\omega_{02}^{2}/\omega_{01}^{2}\right)$ cannot
vanish. Thus, it is not possible to have all real-valued coefficients
in the characteristic polynomials, unless $\gamma_{1}=\gamma_{2}=0$
which corresponds to a lossless circuit. In the following subsection,
we examine the eigenfrequency in a lossless structure to understand
its stability conditions.

\subsubsection{EPD in Lossless Series Circuit\label{subsec:SS}}

\begin{figure}[t]
\centering{}\includegraphics[width=1\columnwidth]{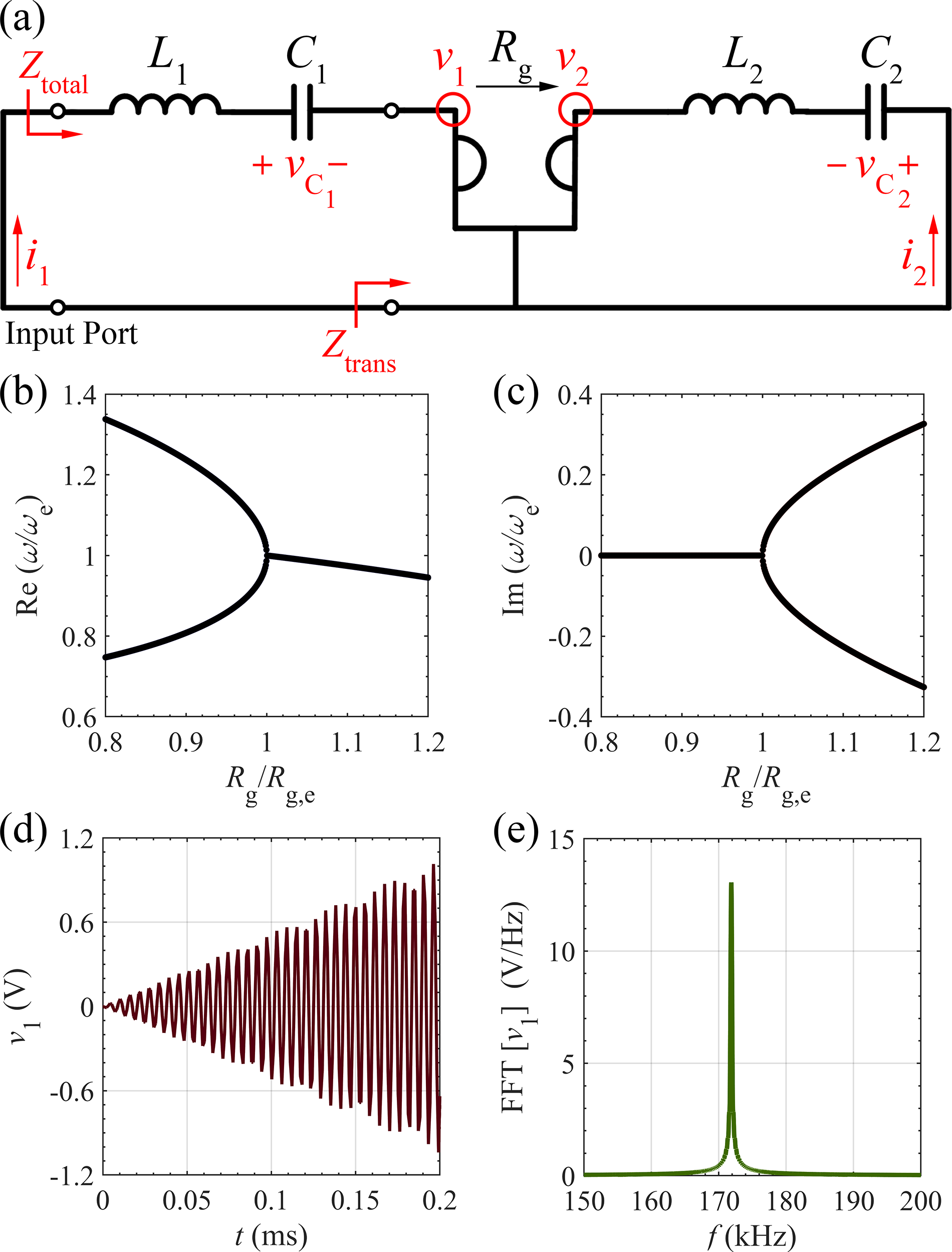}\caption{(a) Series-series configuration: two different LC resonators in series
configuration connected via an ideal gyrator. The right resonator
is made of negative inductance and capacitance. Variation of the (b)
real and (c) imaginary parts of the two eigenfrequencies to a gyration
resistance perturbation. The bifurcation in the real part is observed
for $R_{\mathrm{g}}<R_{\mathrm{g,e}}$. Voltage $v_{1}$ under the
EPD condition in the (d) time domain, and (e) frequency domain. The
frequency domain result is calculated from $150\mathrm{\:kHz}$ to
$200\:\mathrm{kHz}$ with $10^{6}$ samples in the time window between
$0\:\mathrm{ms}$ to $0.2\:\mathrm{ms}$.\label{Fig: LosslessSSCircuitDispersion}}
\end{figure}
To have a real-valued EPD frequency and fulfill the EPD conditions,
we suppose $\gamma_{1}=\gamma_{2}=0$. In this configuration, two
series LC resonators are connected by a gyrator, as illustrated in
Fig. \ref{Fig: LosslessSSCircuitDispersion}(a). We assume that all
components are ideal, and the circuit does not contain any resistance.
By writing down the Kirchhoff voltage law equations in two loops,
we write the eigenvalue problem associated to the circuit equations,
and the characteristic equation is obtained from $\det\left(\underline{\mathbf{M}}-j\omega\underline{\mathbf{I}}\right)=0$,
leading to

\begin{equation}
\omega^{4}-\omega^{2}\left(\omega_{01}^{2}+\omega_{02}^{2}+\frac{R_{\mathrm{g}}}{L_{1}L_{2}}\right)+\omega_{01}^{2}\omega_{02}^{2}=0.\label{eq:SS_Char_Eq}
\end{equation}
In this characteristic equation, $\omega$ is the angular eigenfrequency
of the system. All the $\omega$'s coefficients are real hence $\omega$
and $\omega^{*}$ are both roots of the characteristic equation, where
{*} represents the complex conjugate operation. Moreover, it is a
quadratic equation in $\omega^{2}$; therefore, $\omega$ and $-\omega$
are both solutions. For $R_{\mathrm{g}}=0$, the two resonators are
uncoupled, and the two circuits have two angular eigenfrequency pairs
of $\omega_{1,3}=\pm\omega_{01}$, and $\omega_{2,4}=\pm\omega_{02}$.
We assume that the resonance frequency of each single LC resonator
is real-valued; this happens when inductance and capacitance in the
same resonator have both the same sign. In this case, the component
values on the left side are positive, whereas they are negative on
the right side. We explain the reason for this issue in Appendix \ref{subsec:SS-Components-sign}.
The angular eigenfrequencies (resonance frequencies) in the coupled
circuit are calculated as

\begin{equation}
\omega_{1,3}=\pm\sqrt{a+b},\;\omega_{2,4}=\pm\sqrt{a-b},\label{eq: SSEigenfrequencies}
\end{equation}
where

\begin{equation}
a=\frac{1}{2}\left(\omega_{01}^{2}+\omega_{02}^{2}+\frac{R_{\mathrm{g}}}{L_{1}L_{2}}\right),\label{eq: SSEPD_a}
\end{equation}

\begin{equation}
\begin{array}{c}
b^{2}=a^{2}-\omega_{01}^{2}\omega_{02}^{2}.\end{array}\label{eq: SSEPD_b}
\end{equation}
According to Eq. (\ref{eq: SSEigenfrequencies}), the EPD condition
requires

\begin{equation}
b=0,\label{eq: SS_EPDCondition}
\end{equation}
and the EPD angular frequency is $\omega_{\mathrm{e}}=\pm\sqrt{a}$.
Here, we assume positive values for $a$ in order to have a real EPD
angular frequency and we will only refer to positive values of $\omega_{\mathrm{e}}$
in the following. From Eq. (\ref{eq: SSEPD_b}), the EPD condition
is rewritten as $a^{2}=\omega_{01}^{2}\omega_{02}^{2}.$ Since we
look for real-valued EPD frequencies, $a>0$, and from Eq. (\ref{eq: SSEPD_a})
one has

\begin{equation}
\omega_{01}^{2}+\omega_{02}^{2}-\omega_{\mathrm{gs}}^{2}>0,\label{eq:SS_EPD_Cond_real}
\end{equation}
where it has been convenient to define the equivalent gyrator frequency
$\omega_{\mathrm{gs}}^{2}=-R_{\mathrm{g}}/(L_{1}L_{2})$ for the series-series
configuration (note that $\omega_{\mathrm{gs}}^{2}>0$ because one
inductor is negative). The EPD frequency is calculated by using Eqs.
(\ref{eq: SSEPD_a}), (\ref{eq: SSEPD_b}), and (\ref{eq: SS_EPDCondition})
as

\begin{equation}
\begin{array}{c}
\omega_{\mathrm{e}}=\sqrt{\frac{1}{2}\left(\omega_{01}^{2}+\omega_{02}^{2}-\omega_{gs}^{2}\right)}=\sqrt{\omega_{01}\omega_{02}}.\end{array}\label{eq: SS_EPDFrequency}
\end{equation}

\subsubsection{Dispersion Relation of Lossless and Lossy Series-Series Configurations}

As an example, we explain the required procedure to obtain an EPD
in this configuration by presenting a specific example. Many different
combinations of values for the circuit's components will satisfy the
EPD condition, and here as an example, we assume this set of values
for components: $L_{1}=47\mathrm{\:\mu H}$, $L_{2}=-47\mathrm{\:\mu H}$,
$C_{2}=-47\mathrm{\:nF}$, and $R_{\mathrm{g}}=50\:\Omega$. As mentioned
before, the desired value for the gyration resistance is achieved
by determining the appropriate values for the resistors in the circuit
for the gyrator illustrated in Fig. \ref{Fig: Gyrator}(b). Also,
the capacitance $C_{1}$ is determined by solving the quadratic equation
from the EPD condition in Eq. (\ref{eq: SS_EPDCondition}). There
are two different values of the capacitance $C_{1}$ in the first
resonator that satisfy Eq. (\ref{eq: SS_EPDCondition}), namely $C_{1,\mathrm{e}}=7.05\:\mathrm{nF}$
and $C_{1,\mathrm{e}}=139.16\:\mathrm{nF}$. For the smaller value
($C_{1,\mathrm{e}}=7.05\:\mathrm{nF}$), we obtain a positive value
for $a$ in Eq. (\ref{eq:SS_EPD_Cond_real}), so the EPD frequency
is real. On the contrary, the second value ($C_{1,\mathrm{e}}=139.16\:\mathrm{nF}$)
gives us a negative value for $a$, so the EPD frequency would be
imaginary and we discard it since we investigate a gyrator-based circuit
with real-valued EPD frequency in this paper. In the following, we
select the smaller value for the left resonator capacitance, $C_{1}=7.05\:\mathrm{nF}$.
The results in Figs. \ref{Fig: LosslessSSCircuitDispersion}(b) and
(c) exhibit the two branches of the real and imaginary parts of perturbed
eigenfrequencies obtained from the eigenvalue problem, varying the
gyration resistance $R_{\mathrm{g}}$ in the neighborhood of $R_{\mathrm{g,e}}=50\:\Omega$.
Here, only the two solutions with $\mathrm{Re}\left(\omega\right)>0$
are shown in Figs. \ref{Fig: LosslessSSCircuitDispersion}(b) and
(c). In this example, we obtain $\omega_{\mathrm{e}}=1.08\times10^{6}\:\mathrm{rad/s}$
and the coalesced eigenvalues at EPD are exceedingly sensitive to
perturbations in system parameters.

The time domain simulation results obtained using the Keysight ADS
circuit simulator are illustrated in Figs. \ref{Fig: LosslessSSCircuitDispersion}(d)
and (e). These two plots show the voltage $v_{1}\left(t\right)$ in
the left resonator, and its spectrum, where we put $1\:\mathrm{mV}$
as an initial voltage on $C_{1}$. In the circuit simulator, an ideal
gyrator has been utilized. According to Fig. \ref{Fig: LosslessSSCircuitDispersion}(d),
the voltage grows linearly with increasing time. This important aspect
is peculiar of an EPD, and it is the result of coalescing system eigenvalues
and eigenvectors that also corresponds to a double pole in the system.
A linear growth indicates a second-order EPD in the system. We take
a fast Fourier transform (FFT) of the voltage $v_{1}\left(t\right)$
to show the frequency spectrum, and the calculated result is illustrated
in Fig. \ref{Fig: LosslessSSCircuitDispersion}(e). The observed oscillation
frequency is $f_{\mathrm{o}}=172.05\mathrm{\:kHz}$, which is in good
agreement with the theoretical value $\omega_{\mathrm{e}}/\left(2\pi\right)$
calculated above.

By perturbing the gyration resistance, the operation point moves away
from the EPD. By selecting a lower value for the gyration resistance,
the system has two different real-valued eigenfrequencies. For instance,
we reduce the amount of perturbed parameter by $5\%$ equal to $R_{\mathrm{g}}=47.5\:\Omega<R_{\mathrm{g,e}}=50\:\Omega$.
In the perturbed condition, we do not observe any signal growth in
the system with increasing time. If we consider an additive $5\%$
of perturbation in the gyration resistance, i.e., $R_{\mathrm{g}}=52.5\:\Omega$
>$R_{\mathrm{g,e}}=50\:\Omega$, the imaginary part of the angular
eigenfrequencies is non-zero, and it causes eigensolutions with damping
and growing signals in the system. Since the signal is in the form
of $Q_{n}\varpropto e^{j\omega t}$, the eigenfrequency with negative
imaginary part is associated to an exponentially growing signal.

In lossy circuit, we use the same values as lossless series-series
configuration for the resonators and gyration resistance. In Figs.
\ref{Fig: LossySSCircuitDispersion}(b) and (c), $\gamma_{1}$ is
varied while we assume $\gamma_{2}=0$. In Figs. \ref{Fig: LossySSCircuitDispersion}(d)
and (e), $-\gamma_{2}$ is perturbed while $\gamma_{1}=0$. These
two figures show the real and imaginary parts of eigenfrequencies
when perturbing each resistor individually. The EPD angular frequency
is obtained when $\gamma_{1}=\gamma_{2}=0$, which is the same EPD
frequency as the lossless configuration shown in Section \ref{subsec:SS}.
In Figs. \ref{Fig: LossySSCircuitDispersion}(b)-(e), we observe the
bifurcations of the real and imaginary parts of the eigenfrequencies,
so the circuit is very sensitive to variations in both resistance
values. Angular eigenfrequencies here are complex-valued; it means
that by perturbing $\gamma_{1}$ or $\gamma_{2}$ away from $\gamma_{1}=\gamma_{2}=0$,
the circuit gets unstable; hence it starts to oscillate with the fundamental
frequency associated with the real part of the unstable angular eigenfrequency.
When $\gamma_{1}$ or $\gamma_{2}$ is perturbed from the EPD value,
the oscillation frequency is shifted from the EPD frequency, and it
could be measured for sensing applications. In Figs. \ref{Fig: LossySSCircuitDispersion}(b)-(e),
both conditions $\gamma_{1}>0$ and $-\gamma_{2}>0$ represent losses,
whereas the conditions $\gamma_{1}<0$ and $-\gamma_{2}<0$ represent
gains in the circuit through a negative resistance. In both cases,
by adding either losses or gains, the system is unstable. We observe
more sensitivity when perturbing $R_{2}$, because a small perturbation
in $R_{2}$ results in a larger variation of the eigenfrequencies
than when varying $R_{1}$. Indeed, a wider bifurcation indicates
higher sensitivity.

\subsubsection{Frequency Domain Analysis of The Degenerate Resonance}

\begin{figure}[t]
\centering{}\includegraphics[width=0.7\columnwidth]{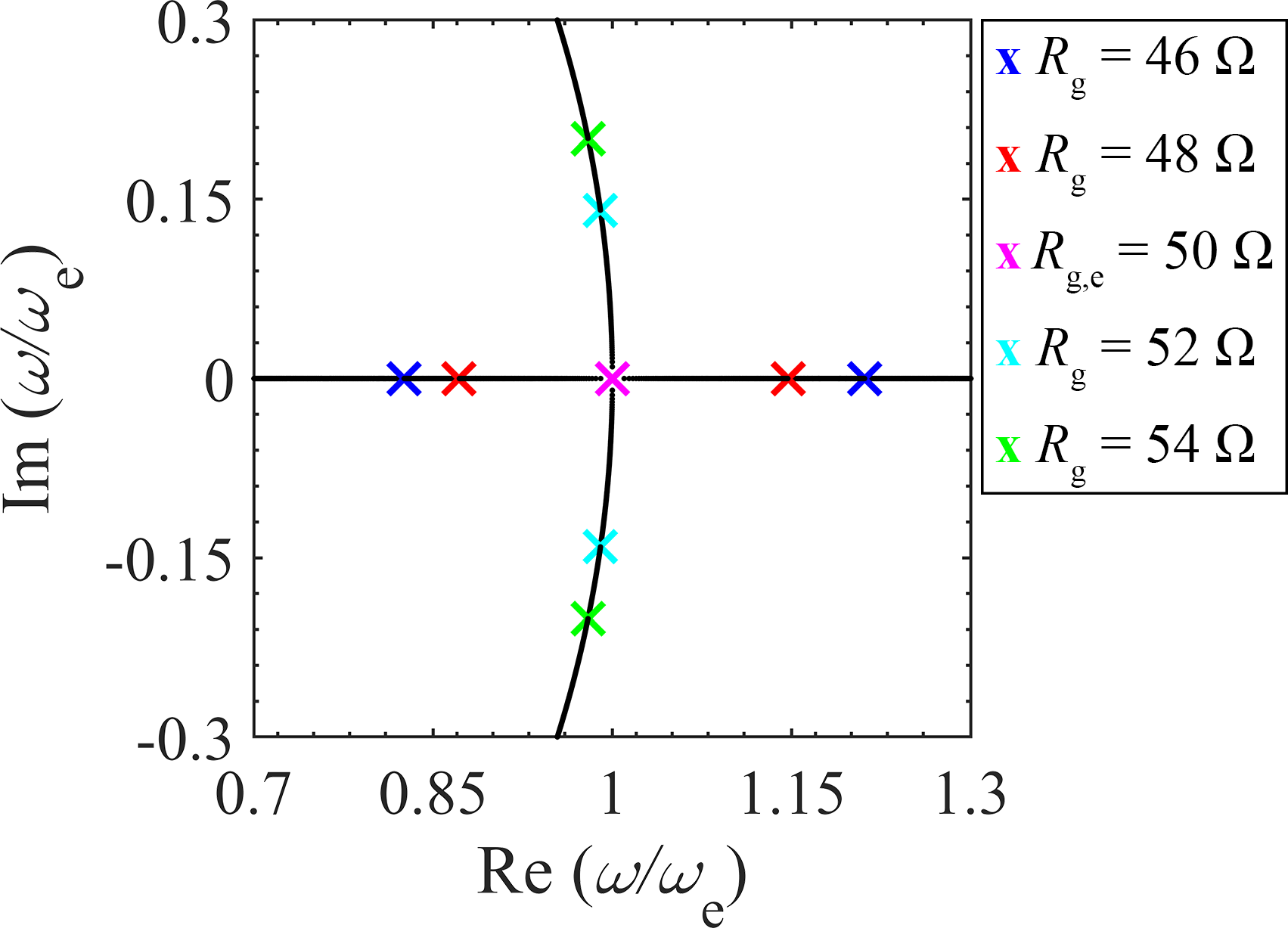}\caption{Root locus of zeros of $Z_{\mathrm{total}}\left(\omega\right)$ of
the series-series configuration when varying the gyration resistance.
The EPD frequency corresponds to a double zero of the impedance $Z_{\mathrm{total}}\left(\omega\right)$.\label{Fig: RootLucasSS}}
\end{figure}

We calculate the total input impedance, $Z_{\mathrm{total}}\left(\omega\right)$
(see Fig. \ref{Fig: LosslessSSCircuitDispersion}), for the series-series
circuit with the same approach discussed in the Section \ref{subsec:Frequency-Domain-Analysis}.
We calculate the transferred impedance on the left side of the circuit
in Fig. \ref{Fig: LosslessSSCircuitDispersion}, that is

\begin{equation}
Z_{\mathrm{trans}}\left(\omega\right)=\frac{R_{\mathrm{g}}^{2}}{Z_{2}}.\label{eq:Ztrans}
\end{equation}
where $Z_{2}\left(\omega\right)=j\omega L_{2}+1/\left(j\omega C_{2}\right)$
is the series impedance on the right side of the circuit. Thus, the
total impedance observed from the input port in this circuit is calculated
by

\begin{equation}
Z_{\mathrm{total}}\left(\omega\right)\triangleq Z_{1}\left(\omega\right)+Z_{\mathrm{trans}}\left(\omega\right)=Z_{1}+\frac{R_{\mathrm{g}}^{2}}{Z_{2}},\label{eq: Ztotal}
\end{equation}
as shown in Fig. \ref{Fig: LosslessSSCircuitDispersion}, where $Z_{1}\left(\omega\right)=j\omega L_{1}+1/\left(j\omega C_{1}\right)$.
The complex-valued resonant frequencies are obtained by imposing $Z_{\mathrm{total}}\left(\omega\right)=0$.
A few steps lead to the $\omega$-zeros given by Eq. (\ref{eq: SSEigenfrequencies}).
Figure \ref{Fig: RootLucasSS} shows the zeros of such total impedance
$Z_{\mathrm{total}}\left(\omega\right)$ for various gyration resistance
values. When considering the EPD gyration resistance $R_{\mathrm{g}}=R_{\mathrm{g,e}}=50\,\Omega$,
one has $Z_{\mathrm{total}}\left(\omega\right)\propto\left(\omega-\omega_{\mathrm{e}}\right)^{2}$,
i.e., the two zeros coincide with the EPD angular frequency $\omega_{\mathrm{e}}$,
that is also the point where the two curves in Fig. \ref{Fig: RootLucasSS}
meet. For gyrator resistances such that $R_{\mathrm{g}}<R_{\mathrm{g,e}}$,
the two resonance angular frequencies are purely real. Instead, for
$R_{\mathrm{g}}>R_{\mathrm{g,e}}$, the two resonance angular frequencies
are complex conjugate, consistent with the result in Fig. \ref{Fig: RootLucasSS}.
In other words, the EPD frequency coincides with double zeros or double
poles of the frequency spectrum, depending on the way the circuit
is described.

\subsection{Components Sign and Simplification of EPD Condition}

\subsubsection{Series-Series Configuration\label{subsec:SS-Components-sign}}

In order to obtain an EPD in the series-series configuration using
Eqs. (\ref{eq: SSEPD_a}), (\ref{eq: SSEPD_b}) and (\ref{eq: SS_EPDCondition})
the following equation must be satisfied:

\begin{equation}
\left(\omega_{01}-\omega_{02}\right)^{2}=\omega_{\mathrm{gs}}^{2}.\label{eq:SS_Components sign}
\end{equation}

We investigate three possible scenarios to satisfy Eq. (\ref{eq:SS_Components sign}).
First, if $\omega_{01}$ and $\omega_{02}$ are pure real, the values
of $L_{1}$ or $L_{2}$ should be negative to have the same sign on
both sides of Eq. (\ref{eq:SS_Components sign}). Thus, one of the
resonators should have a negative inductance to have a pure real $\omega_{01}$
or $\omega_{02}$. Second, if both $\omega_{01}$ and $\omega_{02}$
have imaginary values, the considered values for $L_{1}$ and $L_{2}$
should have the same sign, either positive or negative. When $L_{1}$
and $L_{2}$ are positive, $C_{1}$ and $C_{2}$ should be negative
or vice versa. Finally, if just one of the $\omega_{01}$ or $\omega_{02}$
is imaginary and the other one has a real value, there are no conditions
to obtain an EPD.

\begin{figure}[t]
\centering{}\includegraphics[width=1\columnwidth]{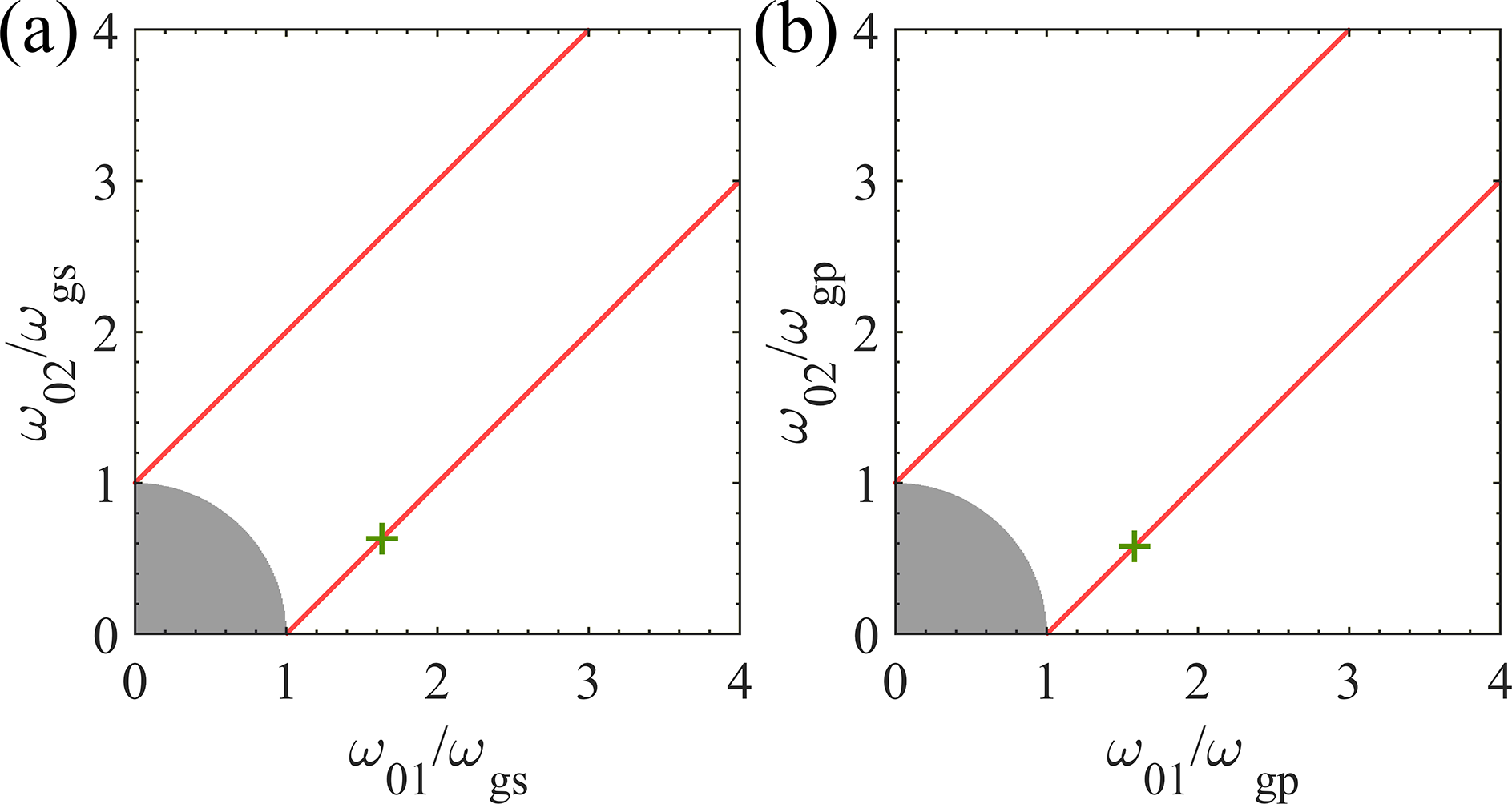}\caption{Possible combinations of $\omega_{01}$ and $\omega_{02}$ to have
a real eigenfrequency in the (a) series-series and (b) parallel-parallel
configurations are shown by the white area, and the gray area represents
complex eigenfrequencies based on (a) Eq. (\ref{eq:SS_EPD_Cond_real})
for series-series configuration and (b) Eq. (\ref{eq:PP_EPD_Cond_real})
for parallel-parallel configuration. Red curves satisfy (a) Eq. (\ref{eq:SS_Components sign})
for series-series configuration and (b) Eq. (\ref{eq:PP_Components sign})
for parallel-parallel configuration and show possible combinations
of $\omega_{01}$ and $\omega_{02}$ that lead to a real EPD frequency.
In the results presented in plot (a), we keep $R_{\mathrm{g}}$, $L_{1}$
and $L_{2}$ fixed as in Section \ref{subsec:SS}, and $C_{1}$ and
$C_{2}$ are varied. In the results presented in plot (b), we keep
$R_{\mathrm{g}}$, $C_{1}$ and $C_{2}$ fixed as in Section \ref{subsec:PP},
and $L_{1}$ and $L_{2}$ are varied.\label{Fig:EPDCoditionsSSPP}}
\end{figure}

To have a real EPD frequency $\omega_{\mathrm{e}}=\pm\sqrt{a}$, $a$
should be positive and this happens when Eq. (\ref{eq:SS_EPD_Cond_real})
is satisfied. The region leading to $a>0$ is represented by the white
area in Fig. \ref{Fig:EPDCoditionsSSPP}(a), whereas the gray area
represents the region with $a<0$. The red curves show different combinations
of $\omega_{01}$ and $\omega_{02}$ which satisfy the EPD condition
of Eq. (\ref{eq:SS_Components sign}), assuming $\omega_{\mathrm{gs}}$
constant. In this figure, $C_{1}=1/\left(\omega_{01}^{2}L_{1}\right)$
and $C_{2}=1/\left(\omega_{02}^{2}L_{2}\right)$ are varied, while
$R_{\mathrm{g}}$, $L_{1}$ and $L_{2}$ are constant. We have shown
only results for positive real values of $\omega_{01}$ and $\omega_{02}$.
The green cross marks the values used for the example provided in
Section \ref{subsec:SS}.

\subsubsection{Parallel-Parallel Configuration\label{subsec:PP-Components-sign}}

In order to get an EPD in the parallel-parallel configuration by using
Eqs. (\ref{eq: PPEPD_a}), (\ref{eq: PPEPD_b}) and (\ref{eq: PP_EPDCondition})
the following condition must be satisfied:

\begin{equation}
\left(\omega_{01}-\omega_{02}\right)^{2}=\omega_{\mathrm{gp}}^{2}.\label{eq:PP_Components sign}
\end{equation}

We consider three different cases for the parallel-parallel configuration
to choose the components' values. First, if $\omega_{01}$ and $\omega_{02}$
are pure real, the values of $C_{1}$ or $C_{2}$ should be negative
to have the same sign on both sides of Eq. (\ref{eq:PP_Components sign}).
Hence, to have a real $\omega_{01}$ and $\omega_{02}$ one resonator
should be made of both negative $C$ and $L$. Second, if both $\omega_{01}$
and $\omega_{02}$ have imaginary values, then $C_{1}$ and $C_{2}$
should have the same sign. Finally, if just one of the $\omega_{01}$
or $\omega_{02}$ is imaginary and the other is real, there is no
condition that leads to an EPD. In this paper, we consider the first
scenario, where both $\omega_{01}$ and $\omega_{02}$ are real.

To have a real EPD frequency $\omega_{\mathrm{e}}=\pm\sqrt{a}$, $a$
should be positive and this occurs when Eq. (\ref{eq:PP_EPD_Cond_real})
is satisfied. The region leading to $a>0$ is represented by the white
area in Fig. \ref{Fig:EPDCoditionsSSPP}(b), whereas the gray area
represents the region with $a<0$. The red curves show different combinations
of $\omega_{01}$ and $\omega_{02}$ which satisfy the EPD condition
of Eq. (\ref{eq:PP_Components sign}), assuming $\omega_{\mathrm{gp}}$
constant. In this figure, $L_{1}=1/\left(\omega_{01}^{2}C_{1}\right)$
and $L_{2}=1/\left(\omega_{02}^{2}C_{2}\right)$ are varied, while
$R_{\mathrm{g}}$, $C_{1}$ and $C_{2}$ are constants. We show only
results for the positive and real values of $\omega_{01}$ and $\omega_{02}$.
The points on the red curves, which are located in the white area,
can be selected to have an EPD with real and positive EPD frequency.
The location marked by the green cross shows the values used for the
example in Subsection \ref{subsec:PP}.

\subsection{The Coefficient of The Leading Term of The Puiseux Series}

Using Eq. (\ref{eq:PuiseuxCoeff1}), we obtain the following expression
for the coefficient of the leading term of the Puiseux series,

\begin{equation}
\alpha_{1}=\sqrt{\frac{\omega_{01}^{2}R_{\mathrm{g}}\left(\omega_{\mathrm{e}}-\omega_{02}^{2}\right)+\frac{\omega_{\mathrm{e}}}{C_{1}C_{2}}}{\frac{1}{C_{1}C_{2}}+R_{\mathrm{g}}\left(\omega_{01}^{2}+\omega_{02}^{2}-6\omega_{\mathrm{e}}\right)}},\label{eq:PuiseuxCoeff1-C}
\end{equation}
when we perturb the capacitance. Instead, when we perturb the inductance,
the coefficient is

\begin{equation}
\alpha_{1}=\sqrt{\frac{\omega_{01}^{2}R_{\mathrm{g}}\left(\omega_{\mathrm{e}}-\omega_{02}^{2}\right)}{\frac{1}{C_{1}C_{2}}+R_{\mathrm{g}}\left(\omega_{01}^{2}+\omega_{02}^{2}-6\omega_{\mathrm{e}}\right)}}.\label{eq:PuiseuxCoeff1-L}
\end{equation}

\subsection{The Impedance Inverter\label{subsec:The-impedance-inverter}}

\begin{figure}[t]
\centering{}\includegraphics[width=0.85\columnwidth]{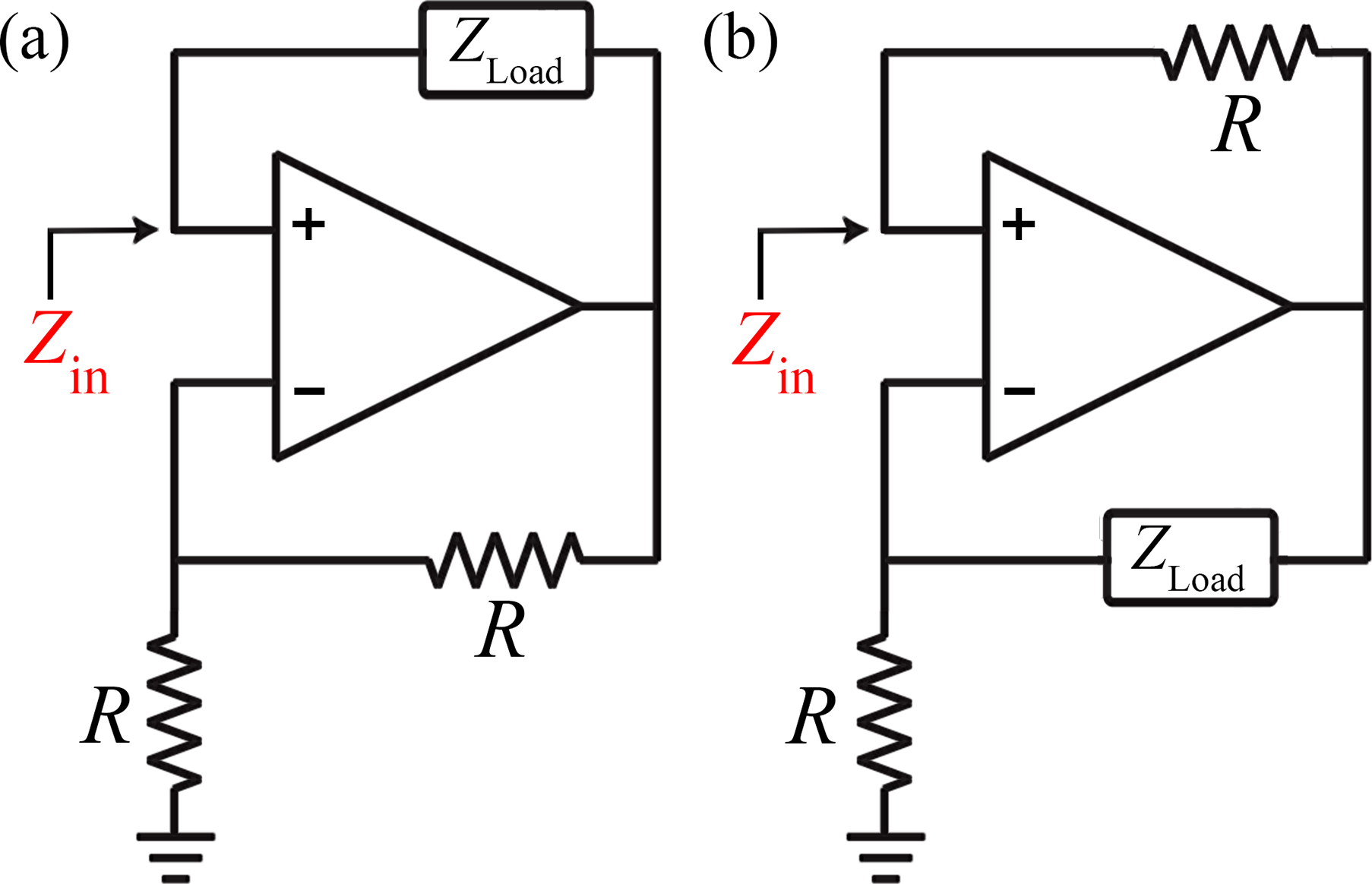}\caption{Negative impedance converter circuit implementations by using an op
amp. (a) Circuit used in the experimental setup where the load is
the parallel between a positive capacitor and a positive inductor.
(b) A negative inductance can be generated using a positive capacitor
(not used in the experimental setup).\label{Fig: Opamp}}
\end{figure}

There are several circuits that can provide for negative capacitances
and inductances needed for the gyrator-based EPD circuits. Two circuits
to obtain negative impedances by using op amps are shown in Fig. \ref{Fig: Opamp}.
The circuit in Fig. \ref{Fig: Opamp}(a) converts the impedance $Z_{\mathrm{Load}}\left(\omega\right)$
to $Z_{\mathrm{in}}\left(\omega\right)=-Z_{\mathrm{Load}}\left(\omega\right)$.
Therefore, when $Z_{\mathrm{Load}}\left(\omega\right)$ in the circuit
in Fig. \ref{Fig: Opamp}(a) is a capacitor in parallel to an inductor,
i.e., $Z_{\mathrm{Load}}\left(\omega\right)=1/\left(j\omega C\right)\parallel\left(j\omega L\right)$,
we obtain $Z_{\mathrm{in}}\left(\omega\right)=-\left(1/\left(j\omega C\right)\parallel(j\omega L)\right)$
at the input port, that corresponds to a negative capacitor in parallel
to a negative inductor. In Sec. \ref{sec:Measurement}, we used this
method to realize negative capacitance and inductance in the measurement.
Figure \ref{Fig: Opamp}(b) shows an alternative way to achieve negative
inductance without an inductor. By using a single capacitor in the
mentioned inverter $Z_{\mathrm{Load}}\left(\omega\right)=1/\left(j\omega C\right)$
resulting in $Z_{\mathrm{in}}\left(\omega\right)=-j\omega R^{2}C$,
hence, desired negative inductance values are achieved with proper
sets of values for $R$ and $C$. Therefore, it is possible to generate
a negative capacitance and a negative inductance by only using capacitive
loads.

\subsection{Implementation of The Gyrator-based Circuit}

\begin{figure}[t]
\centering{}\includegraphics[width=1\columnwidth]{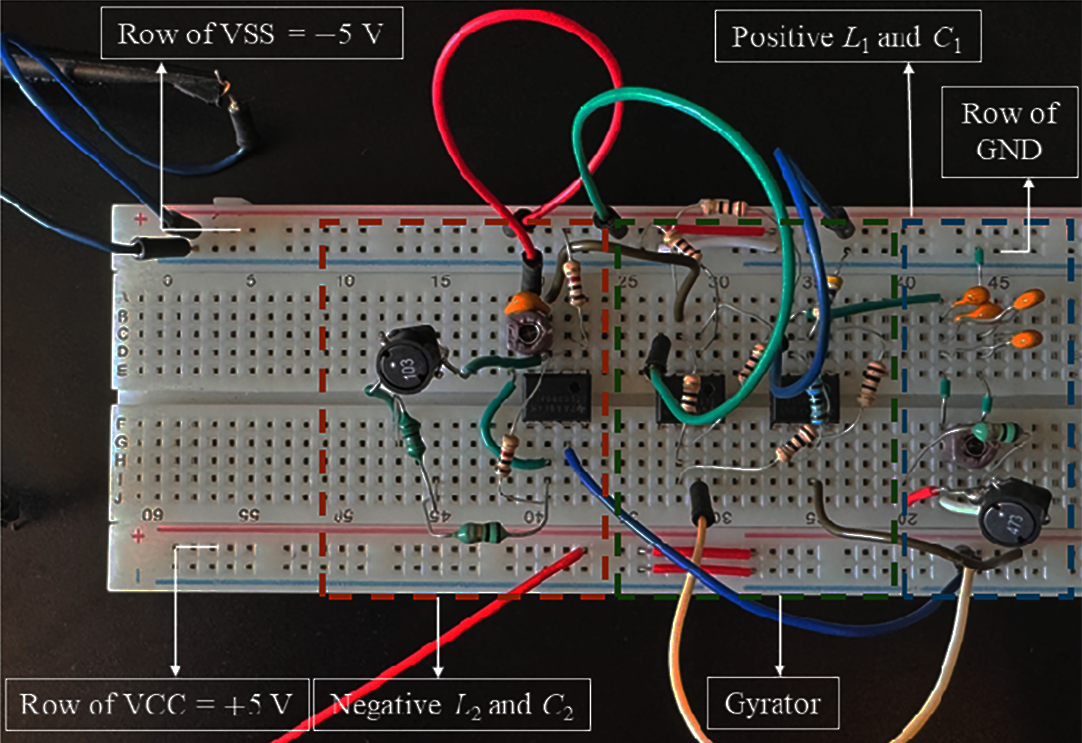}\caption{Assembled gyrator-based oscillator circuit prototype with different
blocks highlighted.\label{fig:Assembled-circuit}}
\end{figure}

Assembled gyrator-based circuit with different blocks highlighted
is shown in Fig. \ref{fig:Assembled-circuit}. The green dashed square
shows the designed gyrator using two op amps, and the red dashed square
shows the inverter circuit to provide a negative inductor in parallel
to a negative capacitor. The circuit also consists of a sensing capacitor
$C_{1}$, where a variable capacitor (FTVOGUE, model Variable Capacitance
Kit) and a series of extra capacitors could be connected in parallel,
as shown in the blue box. To demonstrate the sensitivity of the oscillator's
frequency to perturbations, we perturb the capacitor $C_{1}$ by connecting
pairs of extra $2.5\:\mathrm{nF}$ capacitors in parallel to $C_{1}$.
After each perturbation, the oscillation frequency is measured using
an oscilloscope and a spectrum analyzer for comparison and verification
purposes. Note that on the board, all elements and the DC supply share
a common ground and the VSS ($-5$ V) and VCC ($+5$ V) are connected
to op amps as shown in Fig. \ref{fig:Assembled-circuit}.

\section*{Acknowledgment}

This material is based upon work supported by the National Science
Foundation (NSF) under Grant No. ECCS-1711975 and by Air Force Office
of Scientific Research (AFOSR) under Grant No. FA9550-19-1-0103.

% Generated by IEEEtran.bst, version: 1.14 (2015/08/26)

\end{document}